\documentstyle[amstex,amssymb,preprint,aps,epsf,floats]{revtex}

\begin{document}

\tighten

\preprint{\vbox{
\hbox{UMD-PP-01-004} }} \bigskip \bigskip

\title{Neutrino-Deuteron Scattering in Effective Field Theory at Next-to-Next-to
Leading Order}
\author{Malcolm Butler}
\address{Department of Astronomy and Physics, Saint Mary's University\\
Halifax, NS B3H 3C3 Canada\\
{\tt mbutler@@ap.stmarys.ca}}
\author{Jiunn-Wei Chen}
\address{Department of Physics, University of Maryland, \\
College Park, MD 20742, USA\\
{\tt jwchen@@physics.umd.edu}}
\author{Xinwei Kong}
\address{TRIUMF, 4004 Wesbrook Ave\\
Vancouver BC V6T 2A3 Canada\\
{\tt kong@@alph01.triumf.ca}}
\maketitle
							
\begin{abstract}
We study the four channels associated with neutrino-deuteron breakup
reactions at next-to-next to leading order in effective field theory. We
find that the total cross-section is indeed converging for neutrino energies
up to 20~MeV, and thus our calculations can provide constraints on
theoretical uncertainties for the Sudbury Neutrino Observatory. We stress
the importance of a direct experimental measurement to high precision in at
least one channel, in order to fix an axial two-body counterterm.
\end{abstract}

\vfill\eject

\section{Introduction}

Our understanding of electroweak processes on the deuteron is reaching a
critical juncture. The Sudbury Neutrino Observatory is taking data, and a
thorough understanding of neutrino-deuteron scattering is an important part
of the analysis in that experiment~\cite{SNO,Aardsma}. Of further note,
there is a proposal for a high-precision measurement of the reaction $\nu
_{e}d\rightarrow e^{-}pp$ in the ORLAND experiment~\cite{ORLAND}.

Theorists have made a tremendous effort to understand $\nu (\overline{\nu })$%
-$d$ scattering in a potential model framework~\cite
{ebcc,adnc,HCLee,BKN,TKK,dksno,KN}, with the most recent independent efforts
agreeing within a few percent at low-energy~ \cite{YHH,NSGK}. Given the
ongoing experimental interest in these processes, efforts began to study $%
\nu (\overline{\nu })$-$d$ scattering in the language of effective field
theory (EFT)~\cite{BC1} for neutrino energies below 20~MeV. The EFT work
employed the power-counting scheme of Kaplan, Savage, and Wise\cite{KSW} and
included pions. In working to next-to-leading order (NLO), it was found in
ref.~\cite{BC1} that theoretical uncertainties in the reactions $\nu
d\rightarrow \nu np$, $\overline{\nu }d\rightarrow \overline{\nu }np$, and $%
\overline{\nu }_{e}d\rightarrow e^{+}nn$ were dominated by an unknown axial
two-body counterterm $L_{1,A}$. It was possible to find different values of $%
L_{1,A}$ which provided excellent fits to the potential model calculations
of refs.~\cite{KN,YHH}. Further, we found that the ratio of
charged-current to neutral current (CC/NC) cross-sections was 
insensitive to this counterterm.   This confirmed the insensitivity to 
short-distance physics first discussed by ref.~\cite{YHH}. 
However, it is not clear whether a
power-counting scheme exists which would allow us to extend the theory with
pions to higher-order \cite{CH99,RS99,KS99,FMS}, as would be required to
constrain our
theoretical uncertainties.

In this work, we employ the theory without pions~\cite{CRS} (also see
earlier works~\cite{KSW,KSW96,K97,Cohen97,LM97,L97,vK97,BHvK97}) which has proven so successful in providing
for high-precision calculations for other processes like $np\rightarrow
d\gamma $~\cite{npdgamma,Rupak}. This allows us to consider next-to-next-to
leading order (NNLO) corrections, and to incorporate Coulomb effects
cleanly, following the prescription of Kong and Ravndal~\cite{KR1} with some
generalization. As a result, we present a complete set of calculations for
all four reaction channels including the most-important $pp$ channel. No new
parameters are introduced at this order, so it becomes a
stringent test of the convergence of the calculation and thus on the
theoretical uncertainties. Further, new potential model calculations exist~ 
\cite{NSGK} and a further comparison with those is worthwhile, given that
these new calculations yield different results at threshold to those in
ref.~ \cite{KN}. We continue to emphasize the importance of fixing the axial
counterterm through a direct experimental measurement. This has implications
not only for neutrino-deuteron scattering, but also for $pp\rightarrow
de^{+}\nu _{e}$, $NN\rightarrow NN\overline{\nu }\nu $, and parity violating 
$\vec{e}$-$d$ scattering.

\section{The Lagrangian}

We will briefly review nuclear effective field theory without pions~\cite
{CRS}. The dynamical degrees of freedom are nucleons and non-hadronic
external currents. Massive hadronic excitations such as pions and the delta
are treated as non-dynamical, point interactions whose effects are encoded
in the local operators in the lagrangian. The nucleons are non-relativistic
but with relativistic corrections built in systematically. Nuclear
interaction processes are calculated perturbatively with the small expansion
parameter 
\begin{equation}
Q\equiv \frac{(1/a,\gamma ,p,\left| {\bf q}\right| )}{\Lambda }
\end{equation}
which is the ratio of the light to heavy scales. The light scales include
the inverse S-wave nucleon-nucleon scattering length $1/a(\lesssim 12$ MeV$)$
in the $^{1}S_{0}$ channel, the deuteron binding momentum $\gamma (=45.69$
MeV) in the $^{3}S_{1}$ channel, the magnitude of the nucleon external
momentum $p$ in the two nucleon center-of-mass frame, and the momentum
transfer to the two nucleon system $\left| {\bf q}\right| $. The heavy
scale $\Lambda $ is set by the pion mass $m_{\pi }$. How each term in the
lagrangian scales as powers of $Q$ can be found in ref. \cite{CRS}. It is
the nontrivial renormalization of the strong interaction operators makes the
scaling different from a naive derivative expansion \cite{KSW}.

\subsection{The Effective Lagrangian}

The lagrangian of an effective field theory for nucleons can be
described via 
\begin{equation}
{\cal L}={\cal L}_{1}+{\cal L}_{2}+\cdots \quad ,
\end{equation}
where ${\cal L}_{n}$ contains operators involving $n$ nucleons. Neglecting
for the moment the weak-interaction couplings, we have 
\begin{equation}
{\cal L}_{1}=N^{\dagger }\bigg(iD_{0}+{\frac{{\bf D}^{2}}{2M_{N}}}-\frac{%
D_{0}^{2}}{2M_{N}}\bigg)N\quad ,
\end{equation}
where $N$ is the nucleon field, $M_N$ is the nucleon mass, $D_{0}$ and 
${\bf D}$ are covariant
derivatives and the $D_{0}^{2}$ term is the leading relativistic correction.

The two-nucleon lagrangian needed for a calculation to NNLO is 
\begin{eqnarray}
{\cal L}_{2} &=&-C_{0}^{(\,^{3}S_{1})}(N^{T}P_{i}N)^{\dagger }(N^{T}P_{i}N) 
\nonumber \\
&&+{\frac{C_{2}^{(\,^{3}S_{1})}}{8}}\left[ (N^{T}P_{i}N)^{\dagger }(N^{T}(%
\overleftarrow{{\bf D}}^{2}P_{i}-2\overleftarrow{{\bf D}}\cdot P_{i}%
\overrightarrow{{\bf D}}+P_{i}\overrightarrow{{\bf D}}^{2})N)+h.c.\right]  
\nonumber \\
&&-{\frac{C_{4}^{(\,^{3}S_{1})}}{16}}\bigg(N^{T}\bigg[P_{i}\overrightarrow{%
{\bf D}}^{2}+\overleftarrow{{\bf D}}^{2}P_{i}-2\overleftarrow{{\bf D}}P_{i}%
\overrightarrow{{\bf D}}\bigg]N\bigg)^{\dagger }\bigg(N^{T}\bigg[P_{i}%
\overrightarrow{{\bf D}}^{2}+\overleftarrow{{\bf D}}^{2}P_{i}-2%
\overleftarrow{{\bf D}}P_{i}\overrightarrow{{\bf D}}\bigg]N\bigg)  \nonumber
\\
&&-C_{0}^{(\,^{1}S_{0},i)}(N^{T}\overline{P}_{i}N)^{\dagger }(N^{T}\overline{%
P}_{i}N)  \nonumber \\
&&+{\frac{C_{2}^{(\,^{1}S_{0},i)}}{8}}\left[ (N^{T}\overline{P_{i}}%
N)^{\dagger }(N^{T}(\overleftarrow{{\bf D}}^{2}\overline{P}_{i}-2%
\overleftarrow{{\bf D}}\cdot \overline{P}_{i}\overrightarrow{{\bf D}}+%
\overline{P}_{i}\overrightarrow{{\bf D}}^{2})N)+h.c.\right]   \nonumber \\
&&-{\frac{C_{4}^{(\,^{1}S_{0},i)}}{16}}\bigg(N^{T}\bigg[\overline{P}_{i}%
\overrightarrow{{\bf D}}^{2}+\overleftarrow{{\bf D}}^{2}\overline{P}_{i}-2%
\overleftarrow{{\bf D}}\overline{P}_{i}\overrightarrow{{\bf D}}\bigg]N\bigg)%
^{\dagger }\bigg(N^{T}\bigg[\overline{P}_{i}\overrightarrow{{\bf D}}^{2}+%
\overleftarrow{{\bf D}}^{2}\overline{P}_{i}-2\overleftarrow{{\bf D}}%
\overline{P}_{i}\overrightarrow{{\bf D}}\bigg]N\bigg)\quad ,
\end{eqnarray}
where $P_{i}$ and $\overline{P}_{i}$ are spin-isospin projectors for the $%
^{3}S_{1}$ channel and the $^{1}S_{0}$ channel respectively, 
\begin{eqnarray}
P_{i} &\equiv &\frac{1}{\sqrt{8}}\sigma _{2}\sigma _{i}\tau _{2}\quad ,\quad 
\text{Tr}P_{i}^{\dagger }P_{j}=\frac{1}{2}\delta _{ij}\quad ,  \nonumber \\
\overline{P}_{i} &\equiv &\frac{1}{\sqrt{8}}\sigma _{2}\tau _{2}\tau
_{i}\quad ,\quad \text{Tr}\overline{P}_{i}^{\dagger }\overline{P}_{j}=\frac{1%
}{2}\delta _{ij}\quad ,
\end{eqnarray}
with the $\tau$ matrices act on isospin indices and $\sigma$ matrices on
the spin indices. We incorporate isospin
symmetry breaking in the $C_{2n}^{(\,^{1}S_{0},i)}(n=0,1,...)$ operators,
so that $C_{2n}^{(\,^1S_0,pp)}$, $C_{2n}^{(\,^1S_0,np)}$ and $C_{2n}^{(\,^1S_0,nn)}$ are
different. In both the $^{3}S_{1}$ and $^{1}S_{0}$ channels, the strong coupling
constants $C_{2n}$ have renormalization scale ($\mu $) dependence. These
parameters can be fit to the effective range expansion, as detailed in ref.~ 
\cite{CRS} and as reviewed in Appendix~\ref{appa}.

Relativistic corrections start to contribute to physical quantities at NNLO
and have generic sizes of ${\cal O}(p^{2}/m_{N}^{2})$ of the leading-order
(LO) contribution (see \cite{CRS} for examples). They are suppressed by an
additional factor of $\Lambda ^{2}/m_{N}^{2}$ to other NNLO contributions,
and thus we can neglect them as small (this is verified numerically).

\subsection{Weak Interactions}

The effective lagrangians for charged (CC) and neutral current (NC) weak
interactions are given by 
\begin{equation}
{\cal L}_{{}}^{CC}\ =-{\displaystyle{\frac{G_{F}}{\sqrt{2}}}}l_{+}^{\mu
}J_{\mu }^{-}+h.c.\quad ,
\end{equation}
\begin{equation}
{\cal L}_{{}}^{NC}\ =-{\displaystyle{\frac{G_{F}}{\sqrt{2}}}}l_{Z}^{\mu
}J_{\mu }^{Z}\quad ,
\end{equation}
where the $l_{\mu }$ is the leptonic current and $J_{\mu }$ is the hadronic
current. We have used $G_{F}=1.166\times 10^{-5}$~GeV$^{-2}$. For $\nu $-$d$ and $%
\overline{\nu }$-$d$ scattering, 
\begin{equation}
l_{+}^{\mu }=\overline{\nu }\gamma ^{\mu }(1-\gamma _{5})e\quad ,\quad
l_{Z}^{\mu }=\overline{\nu }\gamma ^{\mu }(1-\gamma _{5})\nu \quad .
\end{equation}
The hadronic currents can be decomposed into vector and axial-vector
contributions 
\begin{eqnarray}
J_{\mu }^{-} &=&V_{\mu }^{-}-%
A_{\mu }^{-}=(V_{\mu }^{1}-A_{\mu }^{1})-i(V_{\mu }^{2}-A_{\mu }^{2})\quad , 
\nonumber \\
J_{\mu }^{Z} &=&-2\sin ^{2}\theta _{W}V_{\mu }^{S}+(1-2\sin ^{2}\theta
_{W})V_{\mu }^{3}-A_{\mu }^{S}-A_{\mu }^{3}\quad ,
\end{eqnarray}
where the superscripts represent isovector components (with $S$ representing
isoscalar terms) and, later, the currents will be labeled by the number of
nucleons involved.

In a NNLO calculation, the electron mass $m_{e}$ contributions to the matrix
elements are counted as higher order (suppressed by a factor of $%
m_{e}^{2}/\gamma ^{2}$), such that $q_{\mu }l^{\mu }=0$
up to NNLO. Similarly, if the neutrino mass $m_{\nu }\neq 0$ but $m_{\nu
}^{2}/\gamma ^{2}\ll 1$, then the massless neutrino treatment we have here
is still applicable. The non-relativistic one-body currents are given by 
\begin{eqnarray}
V_{0}^{(1)} &=&\frac{1}{2}N^{\dagger }(1+\tau _{a})N\quad ,  \nonumber \\
A_{0}^{(1)} &=&-{\frac{i}{2}}N^{\dagger }(\Delta s-g_{A}\tau _{a}){\frac{%
{\bf \sigma }\cdot (\overleftarrow{{\bf \nabla }}-\overrightarrow{{\bf %
\nabla }})}{2M_{N}}}N\quad ,  \nonumber \\
V_{k}^{(1)} &=&{\frac{i}{2}}N^{\dagger }(1+\tau _{a}){\frac{(\overleftarrow{%
\nabla }_{k}-\overrightarrow{\nabla }_{k})}{2M_{N}}}N-N^{\dagger }\bigg(%
\kappa ^{(0)}+\frac{\mu _{s}}{4\sin ^{2}\theta _{W}}+\kappa ^{(1)}\tau _{a}%
\bigg)\epsilon _{kij}{\frac{\sigma _{i}(\overleftarrow{\nabla }_{j}+%
\overrightarrow{\nabla }_{j})}{2M_{N}}}N\quad ,  \nonumber \\
A_{k}^{(1)} &=&-{\frac{1}{2}}N^{\dagger }(\Delta s-g_{A}\tau _{a})\sigma
_{k}N\quad ,  \label{current1}
\end{eqnarray}
where $g_A=1.26$.
Here we have neglected the nucleon vector and axial vector charge radius
contributions. They only contribute at NNLO with about the same size as the
relativistic contributions due to the small momentum transfers being
considered here.

We use the notation $\Delta s$ for the strange quark contribution to the
proton spin 
\begin{equation}
2S_{\mu }\Delta s\equiv \left\langle p\left| \overline{s}\gamma _{\mu
}\gamma _{5}s\right| p\right\rangle \quad  \label{ds}
\end{equation}
with the value \cite{deltas} 
\begin{equation}
\Delta s=-0.17\pm 0.17\>,
\end{equation}
where $S_{\mu }$ is the covariant spin vector. $\kappa ^{(0)}=%
\frac{1}{2}(\kappa _{p}+\kappa _{n})$ and $\kappa ^{(1)}=\frac{1}{2}(\kappa
_{p}-\kappa _{n})$ are the conventional isoscalar and isovector nucleon
magnetic moments in nuclear magnetons, with 
\begin{equation}
\kappa _{p}=2.79285\quad ,\quad \kappa _{n}=-1.91304\quad .
\end{equation}
$\mu _{s}$ is the strange magnetic moment of the proton \cite{KM} 
\begin{gather}
\left\langle p\left| \overline{s}\gamma _{\mu }s\right| p\right\rangle =%
\overline{u}_{p}\bigg(F_1^{s}(q^{2})\gamma_\mu+F_2^{s}(q^{2})\frac{i\sigma
_{\mu
\nu }q^{\nu }}{2M_{N}}\bigg)u_{p}\quad ,  \nonumber \\
G_{M}^{s}(q^2)\equiv F_1^s(q^2)+F_2^s(q^2)\quad,  \nonumber \\
F_1^s(0)=0\quad ,\quad \mu _{s}\equiv G_{M}^{s}(0)\quad
.  \label{mus}
\end{gather}
In ref.~\cite{SAMPLE}, the SAMPLE experiment found
\begin{equation}
G_{M}^{s}(-0.1{\rm GeV}^{2})=0.23\pm 0.37\pm 0.15\pm 0.19\text{ }{\rm n.m.}\>,%
\quad
\end{equation}
extracted from the proton target experiment. However, the newest deuteron target
measurement suggests that the radiative corrections to the axial form factor were
underestimated. Thus the central value of $G_{M}^{s}(-0.1{\rm GeV}^{2})$ could be
40\% smaller \cite{MP}. Theoretical predictions for $\mu _{s}$ range
from -0.8 n.m. to 0.8 n.m..

Finally, there are two-body currents relevant to this process. As mentioned
in ref.~\cite{BC1}, there are axial contributions 
\begin{equation}
A_{k}^{(2)}=L_{1,A}\left( N^{T}P_{k}N\right) ^{\dagger }\left( N^{T}%
\overline{P}_{a}N\right) -2i\varepsilon _{ijk}L_{2,A}(N^{T}P_{i}N)^{\dagger
}(N^{T}P_{j}N)+h.c.\quad ,  \label{2body}
\end{equation}
and analogous vector contributions to the deuteron magnetic properties, as
described in ref.~\cite{CRS,KSW} 
\begin{eqnarray}
V_{k}^{(2)}&=&e\varepsilon _{kij}L_{1}\left( N^{T}P_{i}N\right) ^{\dagger }(%
\overleftarrow{\nabla }_{j}+\overrightarrow{\nabla }_{j})\left( N^{T}%
\overline{P}_{a}N\right)  \nonumber \\
&& -2ieL_{2}(N^{T}P_{i}N)^{\dagger }(\overleftarrow{\nabla }_{i}+%
\overrightarrow{\nabla }_{i})(N^{T}P_{k}N)+h.c.\quad .  \label{2bodyx}
\end{eqnarray}
The parameters $L_{1}$ and $L_{2}$ can be fit from $np\rightarrow d\gamma $
and the deuteron magnetic moment, respectively, and are given by $L_{1}=7.24$%
~fm$^{4}$ and $L_{2}=-0.149$~fm$^{4}$ at $\mu=m_\pi$\cite{CRS}.

\section{$\protect\nu (\bar{\protect\nu})$-$D$ Neutral Current Inelastic
Scattering}

For the inelastic scattering process 
\begin{equation}
\nu +d\rightarrow \nu +n+p,
\end{equation}
the differential cross section can be written in terms of leptonic and
hadronic tensors $l_{\mu \nu }$ and $W_{\mu \nu }$ as 
\begin{equation}
{\displaystyle{d^{2}\sigma  \over d\omega ^{^{\prime }}d\Omega }}%
=%
{\displaystyle{G_{F}^{2}\left| {\bf k}^{\prime }\right|  \over 32\pi ^{2}\left| {\bf k}\right| }}%
S_{1}(\left| {\bf k}^{\prime }\right| )\ l^{\mu \nu }W_{\mu \nu }\quad ,
\label{dsig0}
\end{equation}
where $\omega (\omega ^{\prime })=k_{0}(k_{0}^{\prime })$ represent the
initial(final) neutrino energies. 
\begin{equation}
S_{1}(\left| {\bf k}^{\prime }\right| )=1
\end{equation}
for NC processes. The leptonic tensor is given by 
\begin{equation}
\ l^{\mu \nu }=8(k^{\mu }k^{\nu \prime }+k^{\nu }k^{\mu \prime }-k\cdot
k^{\prime }g^{\mu \nu }+i\varepsilon ^{\mu \nu \rho \sigma }k_{\rho
}k_{\sigma }^{^{\prime }})\quad .  \label{lepten}
\end{equation}
The hadronic tensor is the imaginary part of forward matrix element of the
time order product of two weak current operators. It can be parameterized by
six different structure functions 
\begin{eqnarray}
W_{\mu \nu } &=&%
{\displaystyle{1 \over \pi }}%
\text{Im}\left[ 
\displaystyle\int %
d^{4}xe^{iqx}T\left\langle d(P)\left| J_{\mu }^{Z^{\dagger }}(x)J_{\nu
}^{Z}(0)\right| d(P)\right\rangle \right]  \nonumber \\
&=&-W_{1}g_{\mu \nu }+W_{2}%
{\displaystyle{P_{\mu }P_{\nu } \over M_{d}^{2}}}%
-iW_{3}\varepsilon _{\mu \nu \alpha \beta }%
{\displaystyle{P^{\alpha }q^{\beta } \over M_{d}^{2}}}%
\nonumber \\
&&+W_{4}%
{\displaystyle{q_{\mu }q_{\nu } \over M_{d}^{2}}}%
+W_{5}%
{\displaystyle{(P_{\mu }q_{\nu }+q_{\mu }P_{\nu }) \over M_{d}^{2}}}%
+iW_{6}%
{\displaystyle{(P_{\mu }q_{\nu }-q_{\mu }P_{\nu }) \over M_{d}^{2}}}%
\quad ,
\end{eqnarray}
where the momentum transfer $q_{\mu }=k_{\mu }-k_{\mu }^{\prime }$. We can
easily see that $W_{4},W_{5}$ and $W_{6}$ don't contribute to the
differential cross section because $q_{u}l^{\mu \nu }=0$. In the lab frame
(deuteron rest frame) the differential cross-section simplifies to 
\begin{equation}
{\displaystyle{d^{2}\sigma  \over d\omega ^{^{\prime }}d\Omega }}%
=%
{\displaystyle{G_{F}^{2}\omega ^{\prime }\left| {\bf k}^{\prime }\right|  \over 2\pi ^{2}}}%
S_{1}(\left| {\bf k}^{\prime }\right| )\left[ 2W_{1}\sin ^{2}%
{\displaystyle{\theta  \over 2}}%
+W_{2}\cos ^{2}%
{\displaystyle{\theta  \over 2}}%
-2%
{\displaystyle{(\omega +\omega ^{\prime }) \over M_{d}}}%
W_{3}\sin ^{2}%
{\displaystyle{\theta  \over 2}}%
\right] \quad ,  \label{dsig}
\end{equation}
where\ the $\theta $ is the angle between {\bf k} and {\bf k}$^{^{\prime }}$
and we have used the relation 
\begin{equation}
q^{2}=-4\omega \omega ^{\prime }\sin ^{2}%
{\displaystyle{\theta  \over 2}}%
\quad .
\end{equation}
For $\overline{\nu }d\ \rightarrow \overline{\nu }np$ scattering, the last
terms on the right hand sides of eq.(\ref{lepten}) and (\ref{dsig}) change
sign.

Phase space for this reaction is defined by 
\begin{equation}
\text{Max}\left[ -1,1-%
{\displaystyle{4M_{N}(\nu -B)-\nu ^{2} \over 2\omega \omega ^{\prime }}}%
\right] \leq \cos \theta \leq 1\quad ,  \label{cosnc}
\end{equation}
and 
\begin{equation}
0\leq \omega ^{\prime }\leq \omega -2(M_{N}-\sqrt{M_{N}^{2}-\gamma ^{2}}%
)\quad ,
\end{equation}
where $\nu =\omega -\omega ^{\prime }$, $M_{N}$ is the nucleon mass and $%
B(=2.2245$ MeV) is the deuteron binding energy$.$

\begin{figure}[t]
\centerline{{\epsfxsize=5.0in \epsfbox{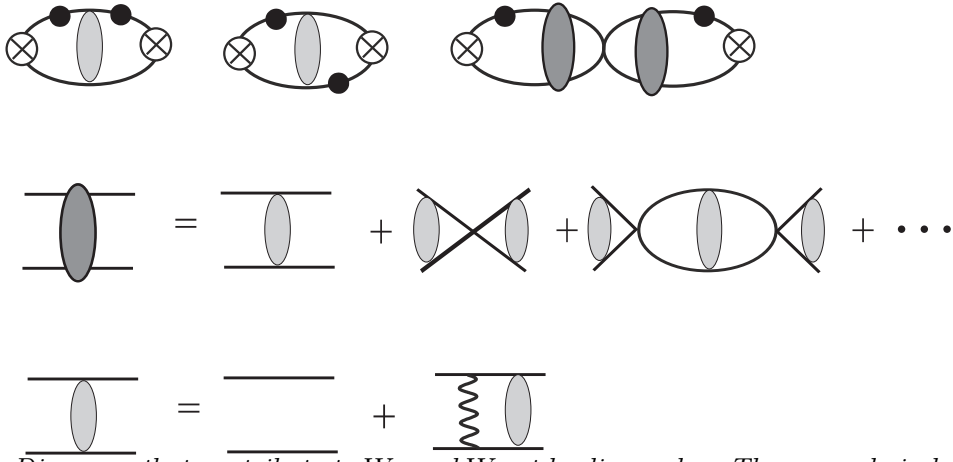}}}
\noindent
\caption{{\it Diagrams that contribute to $W_{0}$ and $W_{1}$ at leading
order. The crossed circles denote operators that create or annihilate two
nucleons with the quantum numbers of the deuteron. The dark gray blobs arise
from scattering corrections involving insertions of $C_{0}$ operators of the
appropriate channel, and from the Coulomb Green's function (the light gray
blob). The wavy line represents the Coulomb interaction, present only in the 
$pp$ final state. Solid lines represent nucleons, and the solid circles
represent insertions of electroweak current operators.}}
\label{LO}
\end{figure}
%

In the deuteron rest frame, 
\begin{eqnarray}
W_{0} &\equiv &W_{00}=W_{2}-W_{1}\quad ,\   \nonumber \\
W_{0i} &=&0\quad ,\   \nonumber \\
W_{ij} &\equiv &\delta _{ij}W_{1}-iW_{3}\varepsilon _{0ijk}%
{\displaystyle{q^{k} \over M_{d}}}%
\quad .
\end{eqnarray}
The structure function can be $Q$ expanded as 
\begin{equation}
W_{\mu \nu }=W_{\mu \nu }^{LO}+W_{\mu \nu }^{NLO}+W_{\mu \nu }^{NNLO}+\cdots
\qquad .
\end{equation}
Now we give the expressions for the structure functions order by order in
the perturbative expansion.

\subsection{Leading Order (LO)}

The LO diagrams in fig.~\ref{LO} give 
\begin{eqnarray}
W_{0}^{LO} &=&\left[
2(C_{V}^{(0)^{2}}+C_{V}^{(1)^{2}})F_{1}+(C_{V}^{(0)^{2}}-C_{V}^{(1)^{2}})F_{2}+4C_{V}^{(0)^{2}}F_{3}^{LO}%
\right] \quad ,  \nonumber \\
W_{1}^{LO} &=&\left[ 2(C_{A}^{(0)^{2}}+C_{A}^{(1)^{2}})F_{1}+%
{\displaystyle{1 \over 3}}%
(C_{A}^{(0)^{2}}-C_{A}^{(1)^{2}})F_{2}+\frac{8}{3}C_{A}^{(0)^{2}}F_{3}^{LO}+%
\frac{4}{3}C_{A}^{(1)^{2}}F_{4}^{LO}\right] \ \quad ,  \nonumber \\
W_{3}^{LO} &=&0\qquad .
\end{eqnarray}
The vector and axial-vector coupling coefficients are given by 
\begin{equation}
\begin{array}{ll}
C_{V}^{(0)}=-\sin ^{2}\vartheta _{W}\quad & ,\quad C_{V}^{(1)}=%
{\displaystyle{1 \over 2}}%
(1-2\sin ^{2}\vartheta _{W})\quad , \\ 
C_{A}^{(0)}=-%
{\displaystyle{1 \over 2}}%
\Delta s\quad & ,\quad C_{A}^{(1)}=%
{\displaystyle{1 \over 2}}%
g_{A}\quad .
\end{array}
\label{couplings}
\end{equation}

The structure and interaction effects are contained in 
\begin{eqnarray}
F_{1} &=&%
{\displaystyle{2M_{N}\gamma \ p \over \pi \left( p^{2}+\gamma ^{2}\right) ^{2}}}%
S_{2}\left( p,|{\bf q}|\right) \quad ,  \nonumber \\
F_{2} &=&%
{\displaystyle{4M_{N}\gamma \ p \over \pi \left( p^{2}+\gamma ^{2}\right) ^{2}}}%
S_{3}\left( p,|{\bf q}|\right) \quad ,  \nonumber \\
F_{3}^{LO} &=&%
{\displaystyle{1\  \over \pi }}%
\text{Im}\left[ B_{0}(p,\left| {\bf q}\right| {\bf )}%
^{2}A_{-1}^{(^{3}S_{1})}(p)\right] \quad ,  \nonumber \\
F_{4}^{LO} &=&F_{3}^{LO}(^{3}S_{1}\rightarrow \ ^{1}S_{0})\ \quad .
\label{fs}
\end{eqnarray}
The magnitude of the relative momentum between the final-state nucleons is $%
2p$, with 
\begin{equation}
p=\sqrt{M_{N}\nu -\gamma ^{2}-\frac{{\bf q}^{2}}{4}+i\epsilon }\quad ,
\end{equation}
where $\epsilon =0^{+}$. For the NC process, 
\begin{eqnarray}
S_{2}\left( p,|{\bf q}|\right) &=&1+\frac{{\bf q}^{2}\left( p^{2}-\gamma
^{2}\right) }{2\left( p^{2}+\gamma ^{2}\right) ^{2}}\quad ,\quad  \nonumber
\\
S_{3}\left( p,|{\bf q}|\right) &=&1-\frac{{\bf q}^{2}\left( p^{2}+3\gamma
^{2}\right) }{6\left( p^{2}+\gamma ^{2}\right) ^{2}}\quad ,  \nonumber \\
B_{0}(p,\left| {\bf q}\right| {\bf )} &=&{\bf -}\sqrt{\frac{\gamma }{2\pi }}%
\frac{M_{N}}{\gamma -ip}\left( 1-\frac{{\bf q}^{2}}{12\left( \gamma
-ip\right) ^{2}}\right) \quad .  \label{bf}
\end{eqnarray}
Note that we have further expanded the ${\bf q}^{2}$ dependence in powers of 
${\bf q}^{2}/\left( p^{2}+\gamma ^{2}\right) $ in order to later obtain analytic
results for the Coulomb contribution in the $pp$ channel. The error introduced is
numerically small ($\ll 1\%$ in total cross section) even though the
neglected terms are formally LO.

The NN scattering amplitude has an expansion 
\begin{equation}
A=A_{-1}+A_{0}+A_{1}+\cdots \qquad ,
\end{equation}
where the subscripts denote the powers in the small expansion parameter $Q$. 
\begin{eqnarray}
A_{-1}^{(^{3}S_{1})}(p) &=&%
{\displaystyle{-4\pi  \over M_{N}}}%
{\displaystyle{1 \over \gamma +ip}}%
\qquad ,  \nonumber \\
A_{0}^{(^{3}S_{1})}(p) &=&%
{\displaystyle{-2\pi  \over M_{N}}}%
{\displaystyle{\rho _{d\ }\left( p^{2}+\gamma ^{2}\right)  \over \left( \gamma +ip\right) ^{2}}}%
\qquad ,  \nonumber \\
A_{1}^{(^{3}S_{1})}(p) &=&%
{\displaystyle{-\pi  \over M_{N}}}%
{\displaystyle{\rho _{d\ }^{2}\left( p^{2}+\gamma ^{2}\right) ^{2} \over \left( \gamma +ip\right) ^{3}}}%
\qquad ,
\end{eqnarray}
where $\rho _{d}=1.764$ fm. And 
\begin{eqnarray}
A_{-1}^{(^{1}S_{0})}(p) &=&%
{\displaystyle{-4\pi  \over M_{N}}}%
{\displaystyle{1 \over %
{\displaystyle{1 \over a^{(^{1}S_{0})}}}+ip}}%
\qquad ,  \nonumber \\
A_{0}^{(^{1}S_{0})}(p) &=&%
{\displaystyle{-2\pi  \over M_{N}}}%
{\displaystyle{r_{0}^{(^{1}S_{0})}p^{2} \over \left( %
{\displaystyle{1 \over a^{(^{1}S_{0})}}}+ip\right) ^{2}}}%
\qquad ,  \nonumber \\
A_{1}^{(^{1}S_{0})}(p) &=&%
{\displaystyle{-\pi  \over M_{N}}}%
{\displaystyle{r_{0}^{(^{1}S_{0})^{2}}p^{4} \over \left( %
{\displaystyle{1 \over a^{(^{1}S_{0})}}}+ip\right) ^{3}}}%
\qquad ,  \label{a1s0}
\end{eqnarray}
where the scattering length $a^{(^{1}S_{0},np)}=-23.7$ fm is known to high
accuracy while the effective range $r_{0}^{(^{1}S_{0},np)}=2.73$ fm has a
2\% uncertainty~\cite{MvO}. This uncertainty is insignificant in this
calculation.

\subsection{Next-to-Leading Order (NLO)}

\begin{figure}[t]
\centerline{{\epsfxsize=3.5 in \epsfbox{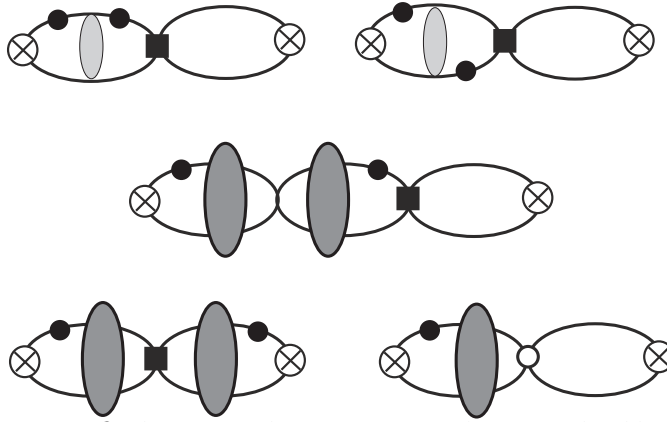}}}  
\noindent
\caption{{\it Diagrams at NLO that contribute to $W_{0}$ and $W_{1}$. The
black square represents an insertion of the $C_{2}$ operator in the $%
^{1}S_{0}$ channel, and $C_{2,-2},\ C_{0,0}$ in the $^{3}S_{1}$ channel. The
white circle represents an insertion of the two-body currents $L_{1,A}$ and $L_{2,A}$. Other
features are as described in fig.~\ref{LO}.}}
\label{NLO}
\end{figure}
%

\begin{figure}[t]
\centerline{{\epsfxsize=4.0in \epsfbox{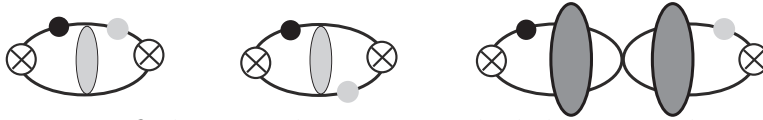}}}
\noindent
\caption{{\it Diagrams at NLO that contribute to $W_{3}$. The light grey
circle represents an insertion of the nucleon magnetic moment operator.
Other features are as described in fig.~\ref{LO}.}}
\label{NLOW3}
\end{figure}
%

At NLO, $W_{0}$ and $W_{1}$ receive contributions from fig.~\ref{NLO} while $%
W_{3}$ receives contributions from fig.~\ref{NLOW3}. 
\begin{eqnarray}
W_{0}^{NLO} &=&\gamma \rho
_{d}W_{0}^{LO}+4C_{V}^{(0)^{2}}F_{3}^{NLO}+G_{1}^{NLO}\quad ,  \nonumber \\
W_{1}^{NLO} &=&\gamma \rho _{d}W_{1}^{LO}+\frac{8}{3}%
C_{A}^{(0)^{2}}F_{3}^{NLO}+\frac{4}{3}C_{A}^{(1)^{2}}F_{4}^{NLO}\
+G_{2}^{NLO}\quad ,  \nonumber \\
W_{3}^{NLO} &=&-2\left[
2(C_{A}^{(0)}C_{M}^{(0)}+C_{A}^{(1)}C_{M}^{(1)})F_{1}+%
{\displaystyle{1 \over 3}}%
(C_{A}^{(0)}C_{M}^{(0)}-C_{A}^{(1)}C_{M}^{(1)})F_{2}+\frac{8}{3}%
C_{A}^{(0)}C_{M}^{(0)}F_{3}^{LO}\right.  \nonumber \\
&&\left. +\frac{4}{3}C_{A}^{(1)}C_{M}^{(1)}F_{4}^{LO}\right] \ \quad ,
\end{eqnarray}
where 
\begin{eqnarray}
F_{3}^{NLO} &=&F_{3}^{LO}(A_{-1}^{(^{3}S_{1})}\rightarrow
A_{0}^{(^{3}S_{1})})\qquad ,  \nonumber \\
F_{4}^{NLO} &=&F_{4}^{LO}(A_{-1}^{(^{1}S_{0})}\rightarrow
A_{0}^{(^{1}S_{0})})\qquad ,
\end{eqnarray}
and
\begin{eqnarray}
G_{1}^{NLO} &=&C_{V}^{(0)^{2}}%
{\displaystyle{2M_{N}\rho _{d} \over \pi }}%
\mathop{\rm Im}%
\left[ \sqrt{\frac{2\gamma }{\pi }}B_{0}(p,\left| {\bf q}\right| {\bf )}\
A_{-1}^{(^{3}S_{1})}(p)\right] \quad ,  \nonumber \\
G_{2}^{NLO} &=&-%
{\displaystyle{M_{N} \over 3\pi ^{2}}}%
\mathop{\rm Im}%
\left[ \sqrt{\frac{\gamma }{2\pi }}B_{0}(p,\left| {\bf q}\right| {\bf )}%
\left( C_{A}^{(1)}\widetilde{L}_{1,A}A_{-1}^{(^{1}S_{0})}(p)+4C_{A}^{(0)}%
\widetilde{L}_{2,A}A_{-1}^{(^{3}S_{1})}(p)\right) \right] \qquad .
\end{eqnarray}
Note that $G_{1}^{LO}=G_{2}^{LO}=0$. The $\widetilde{L}_{A}$'s are
renormalization scale $\mu $-independent quantities defined as 
\begin{eqnarray}
\widetilde{L}_{1,A} &=&-\frac{4\pi \left( \mu -\gamma \right) }{%
M_{N}C_{0}^{(^{1}S_{0})}}\left[ L_{1,A}-2\pi C_{A}^{(1)}\left( \frac{M_{N}}{%
2\pi }C_{2}^{(^{1}S_{0})}+\frac{\rho _{d}}{\left( \mu -\gamma \right) ^{2}}%
\right) \right] \quad ,  \label{LA} \\
\widetilde{L}_{2,A} &=&\left( \mu -\gamma \right) ^{2}L_{2,A}-2\pi
C_{A}^{(0)}\rho _{d}\quad .
\end{eqnarray}
The expressions for $C_{0}^{(^{1}S_{0})}$ and $C_{2}^{(^{1}S_{0})}$ can be
found in Appendix~\ref{appa}. Through $W_{3}$, we become sensitive to weak
magnetism at NLO, with coupling coefficients given by 
\begin{equation}
\begin{array}{ll}
C_{M}^{(0)}=-2\sin ^{2}\vartheta _{W}\kappa ^{(0)}-\frac{1}{2}\mu _{S} & 
,\quad C_{M}^{(1)}=(1-2\sin ^{2}\vartheta _{W})\kappa ^{(1)}\quad .
\end{array}
\end{equation}

\begin{figure}[t]
\centerline{{\epsfxsize=4 in \epsfbox{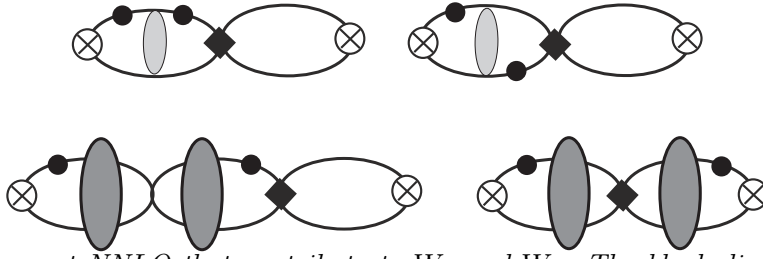}}} 
\noindent
\caption{{\it Diagrams at NNLO that contribute to $W_{0}$ and $W_{1}$. The
black diamond represents an insertion of the $C_{4}$ operator in the $%
^{1}S_{0}$ channel, and $C_{4,-3},\ C_{2,-1},\ C_{0,1}$ in the $^{3}S_{1}$
channel. Other features are as described in fig.~\ref{LO}.}}
\label{NNLO1}
\end{figure}
%

\begin{figure}[t]
\centerline{{\epsfxsize=5.5in \epsfbox{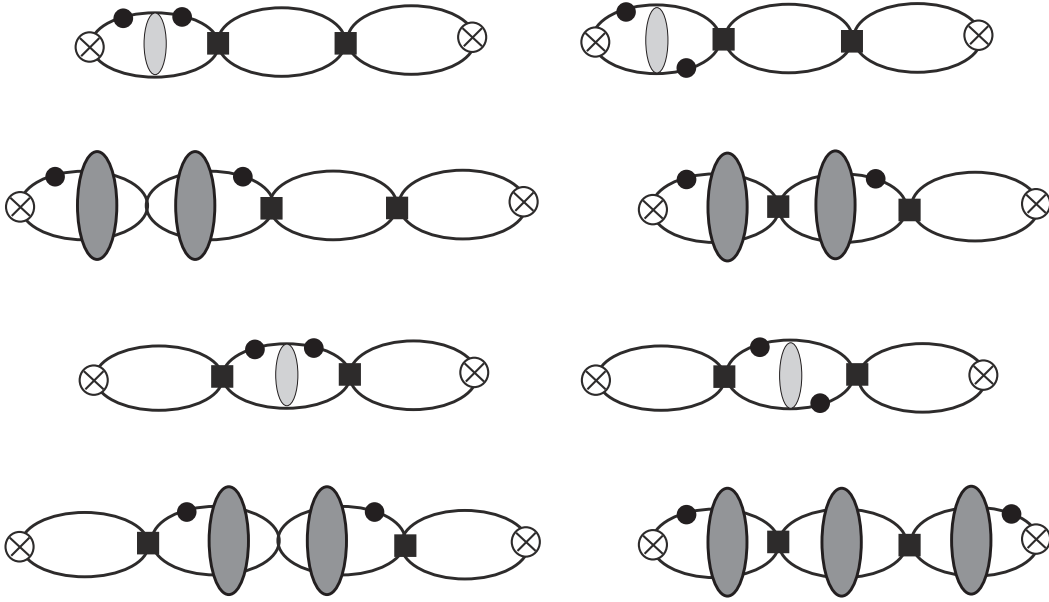}}}
\noindent
\caption{{\it Diagrams at NNLO that contribute to $W_{0}$ and $W_{1}$
through two insertions (black squares) of the $C_{2}$ operator in the $%
^{1}S_{0}$ channel, and $C_{2,-2},\ C_{1,0}$ in the $^{3}S_{1}$ channel.
Other features are as described in fig.~\ref{LO}.}}
\label{NNLO2}
\end{figure}
%
\begin{figure}[t]
\centerline{{\epsfxsize=5.0in \epsfbox{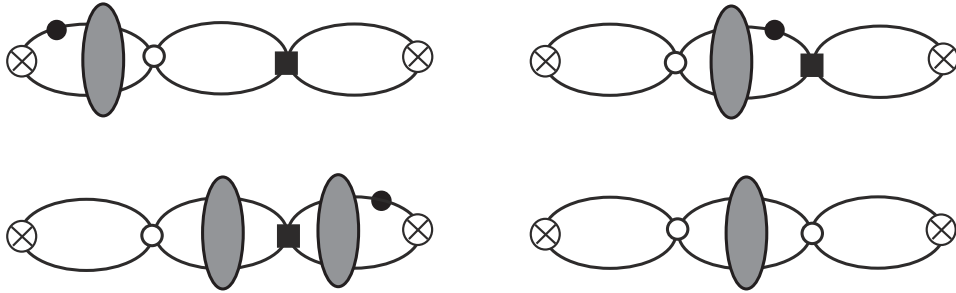}}}
\noindent
\caption{{\it Diagrams at NNLO that contribute to $W_{0}$ and $W_{1}$
through mixed insertions of the $L_{1,A}$ and $L_{2,A}$ operators (white
circle) and $C_{2}$ operators (black square). Other features are as
described in fig.~\ref{LO}.}}
\label{NNLO3}
\end{figure}
%

\begin{figure}[t]
\centerline{{\epsfxsize=4.5in \epsfbox{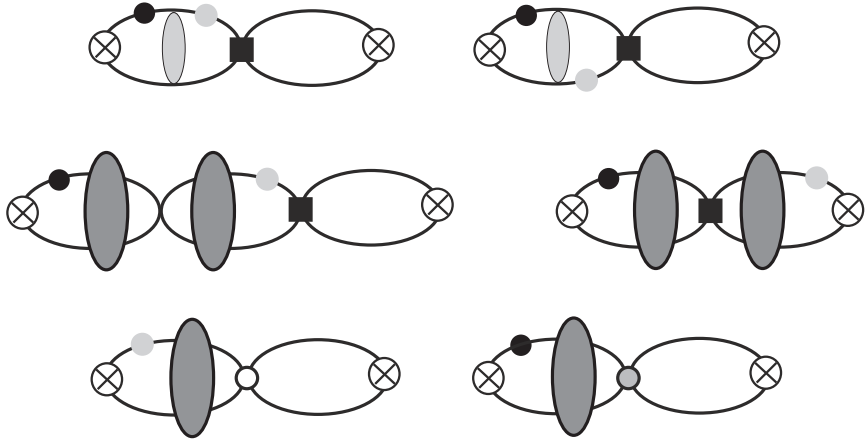}}}
\noindent
\caption{{\it Diagrams at NNLO that contribute to $W_{3}$. The filled circle
represents an insertion of the two-body current $L_1$ or $L_2$. Other
features are as described in figs.~\ref{LO} and \ref{NLO}. }}
\label{NNLOW3}
\end{figure}
%

\subsection{Next-to-Next-to-Leading Order (NNLO)}

Finally, at NNLO, $W_{0}$ and $W_{1}$ receive contributions from figs.~\ref
{NNLO1}-\ref{NNLO3} while $W_{3}$ receives contributions from fig.~\ref
{NNLOW3}. 
\begin{eqnarray}
W_{0}^{NNLO} &=&\gamma \rho
_{d}W_{0}^{NLO}+4C_{V}^{(0)^{2}}F_{3}^{NNLO}+G_{1}^{NNLO}\quad ,  \nonumber
\\
W_{1}^{NNLO} &=&\gamma \rho _{d}W_{1}^{NLO}+\frac{8}{3}%
C_{A}^{(0)^{2}}F_{3}^{NNLO}+\frac{4}{3}C_{A}^{(1)^{2}}F_{4}^{NNLO}\
+G_{2}^{NNLO}\quad ,  \nonumber \\
W_{3}^{NNLO} &=&\gamma \rho _{d}W_{3}^{NLO}-2\left( \frac{8}{3}%
C_{A}^{(0)}C_{M}^{(0)}F_{3}^{NLO}+\frac{4}{3}%
C_{A}^{(1)}C_{M}^{(1)}F_{4}^{NLO}+G_{3}^{NLO}\right) \qquad ,
\end{eqnarray}
where 
\begin{eqnarray}
F_{3}^{NNLO} &=&F_{3}^{LO}(A_{-1}^{(^{3}S_{1})}\rightarrow
A_{1}^{(^{3}S_{1})})\qquad ,  \nonumber \\
F_{4}^{NNLO} &=&F_{4}^{LO}(A_{-1}^{(^{1}S_{0})}\rightarrow
A_{1}^{(^{1}S_{0})})\qquad .
\end{eqnarray}
\begin{eqnarray}
G_{1}^{NNLO} &=&G_{1}^{NLO}(A_{-1}^{(^{3}S_{1})}\rightarrow
A_{0}^{(^{3}S_{1})})+C_{V}^{(0)^{2}}%
{\displaystyle{\gamma M_{N}^{2}\rho _{d}^{2} \over 2\pi ^{2}}}%
\mathop{\rm Im}%
\left[ A_{-1}^{(^{3}S_{1})}(p)\right] \quad ,  \nonumber \\
G_{2}^{NNLO} &=&G_{2}^{NLO}(A_{-1}^{(^{3}S_{1})}\rightarrow
A_{0}^{(^{3}S_{1})},A_{-1}^{(^{1}S_{0})}\rightarrow A_{0}^{(^{1}S_{0})}) 
\nonumber \\
&&+%
{\displaystyle{M_{N}^{2}\gamma  \over 96\pi ^{4}}}%
\mathop{\rm Im}%
\left[ \left( \widetilde{L}_{1,A}^{2}A_{-1}^{(^{1}S_{0})}(p)+8\widetilde{L}%
_{2,A}^{2}A_{-1}^{(^{3}S_{1})}(p)\right) \right] \quad , \\
G_{3}^{NLO} &=&-%
{\displaystyle{M_{N} \over 6\pi ^{2}}}%
\mathop{\rm Im}%
\left\{ \sqrt{\frac{\gamma }{2\pi }}B_{0}(p,\left| {\bf q}\right| {\bf )}%
\left[ \left( C_{M}^{(1)}\widetilde{L}_{1,A}+C_{A}^{(1)}\widetilde{L}_{1,M}\
\right) A_{-1}^{(^{1}S_{0})}(p)\right. \right.  \nonumber \\
&&\left. \left. +4\left( C_{M}^{(0)}\widetilde{L}_{2,A}+C_{A}^{(0)}%
\widetilde{L}_{2,M}\right) A_{-1}^{(^{3}S_{1})}(p)\right] \right\} \>. \nonumber
\end{eqnarray}
Note that $G_{3}^{LO}=G_{3}^{NLO}=0$. The $\widetilde{L}_{M}$'s are $\mu $
independent quantities defined as 
\begin{eqnarray}
\widetilde{L}_{1,M} &=&-\frac{8\pi \left( \mu -\gamma \right) }{%
M_{N}C_{0}^{(^{1}S_{0})}}\left[ (1-2\sin ^{2}\vartheta _{W})M_{N}L_{1}-\pi
C_{M}^{(1)}\left( \frac{M_{N}}{2\pi }C_{2}^{(^{1}S_{0})}+\frac{\rho _{d}}{%
\left( \mu -\gamma \right) ^{2}}\right) \right] \quad , \label{LM}
\\
\widetilde{L}_{2,M} &=&\ -4\sin ^{2}\vartheta _{W}\left( \mu -\gamma \right)
^{2}M_{N}L_{2}-2\pi C_{M}^{(0)}\rho _{d}\quad .  
\end{eqnarray}

At NNLO, we should mention the effects of other partial waves, beyond $S$%
-wave. The $P$-wave {\it NN} rescattering does not contribute at NNLO. The $%
D $-wave would contribute to $W_{4}$ at NNLO, but this structure function
does not contribute to the cross-section, so we can neglect $D$-wave initial
and final states, also.

A summary of the expressions in both this and the next section, order-by-order,
can be found in Appendix~\ref{appb}.

\section{$\protect\nu (\bar{\protect\nu})$-$D$ Charged Current Inelastic
Scattering}

For CC processes, a few inputs change from their NC values and there are
effects of electron/positron mass to consider. For $\overline{\nu }%
d\rightarrow e^{+}nn$, the sign of the last term of eq.\thinspace (\ref{dsig}%
) is +. $S_{1}$, $S_{2}$, $S_{3}$, and $B_{0}$ in eqs.\thinspace (\ref{dsig0}%
), (\ref{dsig}) and (\ref{fs}) still take the same functional forms as for NC
channels. The phase space is modified, due to the positron mass $m_{e}$ and the
neutron--proton mass splitting $\delta m=m_{n}-m_{p}$, to be 
\begin{gather}
\text{Max}\left[ -1,-{\displaystyle{\frac{4M_{N}(\nu -B-\delta m)-\omega
^{2}-\omega ^{\prime 2}+m_{e}^{2}}{2\omega \sqrt{\omega ^{\prime 2}-m_{e}^{2}%
}\ }}}\right] \leq \cos \theta \leq 1\quad ,  \nonumber \\
m_{e}\leq \omega ^{\prime }\leq \omega -2\bigg(M_{N}-\sqrt{M_{N}^{2}-M_{N}(B+%
\delta m)} \bigg)  \label{ps2}
\end{gather}
In principle, there are also electron mass corrections to eq.~(\ref{dsig}), but
these are only important close to threshold and that region will not be
probed by SNO.

For the most part, however, the primary difference between the neutral
current and charged current cases is the fact that the charged current
processes are purely isovector. As a result, the charged current results can
be obtained from the neutral current structure factors with the
substitutions: 
\begin{gather}
C_{V}^{(0)}=0\quad ,\quad C_{V}^{(1)}={{%
{\displaystyle{\left| V_{ud}\right|  \over \sqrt{2}}}%
}}\quad ,\quad  \nonumber \\
C_{A}^{(0)}=0\quad ,\quad C_{A}^{(1)}={{%
{\displaystyle{\left| V_{ud}\right|  \over \sqrt{2}}}%
}}g_{A}\quad ,  \nonumber \\
C_{M}^{(0)}=0\quad ,\quad C_{M}^{(1)}=\sqrt{2}\left| V_{ud}\right| \kappa
^{(1)}\quad ,  \nonumber \\
L_{1,A}\rightarrow \sqrt{2}L_{1,A}\left| V_{ud}\right| \quad ,\quad
L_{2,A}=0\quad ,  \nonumber \\
(1-2\sin ^{2}\vartheta _{w})L_{1}\rightarrow \sqrt{2}|V_{ud}|L_{1}\quad
L_{2}=0  \nonumber \\
\quad \nu \rightarrow \nu -\delta m\quad ,  \label{ccp}
\end{gather}
where we use $\left| V_{ud}\right| =0.975$ for this CKM matrix element. The $%
nn$ scattering amplitude still has the same form as eq.~(\ref{a1s0}), as
do eqs.~(\ref{LA}) and (\ref{LM}), but with different effective range
parameters. We have used $a^{(^{1}S_{0},nn)}=-18.5$ fm and $%
r_{0}^{(^{1}S_{0},nn)}=2.80$ fm. Ref. \cite{MvO} indicates the uncertainty
in $a^{(^{1}S_{0},nn)}$ is at the few percent level. It is important to note
that a 2\% uncertainty in $a^{(^{1}S_{0},nn)}$ will change the $\nu $-$d$
breakup cross section at threshold by 3-4\%.

For $\nu _{e}d\rightarrow e^{-}pp$, electromagnetic corrections in the final
state are important. But instead of solving a three-body problem, we can
factor out the Coulomb interaction between the electron and two protons by
the Sommerfeld factor 
\begin{eqnarray}
S_{1}(\left| {\bf k}^{\prime }\right| ) &=&\frac{2\pi \eta _{e}}{1-e^{-2\pi
\eta _{e}}\ }\ \quad ,  \label{Sommerfeld} \\
\eta _{e} &=&\frac{2\alpha \omega ^{\prime }}{\left| {\bf k}^{\prime
}\right| }\quad .
\end{eqnarray}
This approximation is valid because the strength of a single-photon exchange
between two particles with relative velocity $v$ scales as $\alpha /v$
(Coulomb effect). The effect becomes sizable only when the photon exchange
becomes nonperturbative, i.e. $v\lesssim \alpha $. For an electron, this
velocity corresponds to a wavelength much longer than the size of the two
proton system, thus eq.~(\ref{Sommerfeld}) is justified.

For the proton-proton electromagnetic interaction, the Coulomb contribution
is enhanced by a factor of $1/v$ and dominates over other short-distance
photon exchange processes. We explicitly compute the long-distance Coulomb
contribution and encode the short-distance photon effects into the local
operators which are then fit to data. In this way we naturally incorporate
all the isospin symmetry breaking effects in the calculation, except for the
unknown counterterm $L_{1,A}$ contribution. Since the $L_{1,A}$ operator
only encodes short-distance physics, the only contribution to
isospin symmetry breaking is through hard photons, and thus the effect is of
 $\sim O(\alpha )$ and thus negligible.
The other symmetry breaking effects on $L_{1,A}$ will be estimated later.

For $\nu _{e}d\rightarrow e^{-}pp$ the sign of the last term of
eq.\thinspace (\ref{dsig}) is negative($-$). The effects of the Coulomb
interaction are encoded in the $S$'s and $B_{0}$ as 
\begin{eqnarray}
S_{2}\left( p,|{\bf q}|\right) &=&\left( 1+\frac{{\bf q}^{2}\left(
p^{2}-\gamma ^{2}-2\eta p\gamma \right) }{2\left( p^{2}+\gamma ^{2}\right)
^{2}}\right) \left( 
{\displaystyle{2\pi \eta  \over \ e^{2\pi \eta }-1\ }}%
\right) e^{4\eta \tan ^{-1}\left( \frac{p}{\gamma }\right) }\quad ,\quad 
\nonumber \\
S_{3}\left( p,|{\bf q}|\right) &=&\left( 1-\frac{{\bf q}^{2}\left(
p^{2}+3\gamma ^{2}+6\eta p\gamma \right) }{6\left( p^{2}+\gamma ^{2}\right)
^{2}}\right) \left( 
{\displaystyle{2\pi \eta  \over \ e^{2\pi \eta }-1\ }}%
\right) e^{4\eta \tan ^{-1}\left( \frac{p}{\gamma }\right) }\quad , 
\nonumber \\
\eta &=&\frac{\alpha M_{N}}{2p}\quad .
\end{eqnarray}
\begin{eqnarray}
B_{0}(p,\left| {\bf q}\right| {\bf )} &=&M_{N}\int \frac{d^{3}k}{\left( 2\pi
\right) ^{3}}\frac{\sqrt{8\pi \gamma }}{k^{2}+\gamma ^{2}}%
{\displaystyle{2\pi \eta _{k} \over e^{2\pi \eta _{k}}-1\ }}%
\frac{e^{2\eta _{k}\tan ^{-1}\left( \frac{k}{\gamma }\right) }}{%
p^{2}-k^{2}+i\epsilon }  \nonumber \\
&&\qquad \qquad \qquad \qquad \left[ 1+\frac{{\bf q}^{2}\left( \left(
1-2\eta _{k}^{2}\right) k^{2}-6\eta _{k}k\gamma -3\gamma ^{2}\right) }{%
12\left( q^{2}+\gamma ^{2}\right) ^{2}}\right] \quad ,  \nonumber \\
\eta _{k} &=&\frac{\alpha M_{N}}{2k}\quad .
\end{eqnarray}

Finally, for the $pp$ channel, eq.~(\ref{a1s0}) must be replaced by (see
Appendix~\ref{appa}) 
\begin{eqnarray}
A_{-1}^{(^{1}S_{0},pp)}(p) &=&%
{\displaystyle{-4\pi  \over M_{N}}}%
{\displaystyle{1 \over %
{\displaystyle{1 \over a^{(^{1}S_{0},pp)}}}+\alpha M_{N}H(\eta )}}%
\qquad ,  \nonumber \\
A_{0}^{(^{1}S_{0},pp)}(p) &=&%
{\displaystyle{-2\pi  \over M_{N}}}%
{\displaystyle{r_{0}^{(^{1}S_{0},pp)}p^{2} \over \left( %
{\displaystyle{1 \over a^{(^{1}S_{0},pp)}}}+\alpha M_{N}H(\eta )\right) ^{2}}}%
\qquad ,  \nonumber \\
A_{1}^{(^{1}S_{0},pp)}(p) &=&%
{\displaystyle{-\pi  \over M_{N}}}%
{\displaystyle{r_{0}^{(^{1}S_{0},pp)^{2}}p^{4} \over \left( %
{\displaystyle{1 \over a^{(^{1}S_{0},pp)}}}+\alpha M_{N}H(\eta )\right) ^{3}}}%
\qquad ,  \label{app}
\end{eqnarray}
where 
\begin{equation}
H(\eta )=\psi (i\eta )+\frac{1}{2i\eta }-\ln (i\eta )
\end{equation}
with $\psi $ the logarithmic derivative of the $\Gamma $-function. Note that 
$A^{(^{1}S_{0},pp)}$ in eq.\thinspace (\ref{app}) is $not$ the full $pp$
scattering amplitude. It is the $pp$ scattering amplitude with the
pure Coulomb phase
shift removed, as discussed in Appendix~\ref{appa}. We have
used $a^{(^{1}S_{0},pp)}=-7.82$ fm and $r_{0}^{(^{1}S_{0},pp)}=2.79$ fm.
These are known to high accuracy. Furthermore, the values for $%
C_{0}^{(^{1}S_{0})}$ and $C_{2}^{(^{1}S_{0})}$ used in eqs.~(\ref{LA}) and (%
\ref{LM}) should be replaced by $C_{0}^{(^{1}S_{0},pp)}$ and $%
C_{2,-2}^{(^{1}S_{0},pp)}$ as discussed in Appendix~\ref{appa}.

Finally, modifications to the phase space and parameters in eqs.\thinspace (%
\ref{ps2}) and (\ref{ccp}) correspond to simply changing the sign of $\delta
m,$%
\begin{equation}
\delta m\rightarrow -\delta m\quad .
\end{equation}

We note again that the expressions from this section, and the previous one,
are summarised in Appendix~\ref{appb}.

\section{Results}

\subsection{Unknown Parameters}

As discussed extensively in ref.\cite{BC1}, contributions from $\Delta s$, $%
\mu _{s}$ and $L_{2,A}$ which are not well constrained are, in fact,
negligible ($\ll 1\%$) due to quasi-orthogonality between initial and final
states in the $^{3}S_{1}$ channel. Thus, up to NNLO the axial two-body
counter term $L_{1,A}$ is the only unknown parameter contributing to each
breakup channel. To estimate the effect of isospin-symmetry breaking on $%
L_{1,A}$, we consider how much $L_{1,A}$ must vary in order for $\widetilde{L%
}_{1,A}$ (defined in eq.(\ref{LA})) to take on a universal value. This
assumes that the symmetry breaking effects in $C_{0}^{(^{1}S_{0})}$, $%
C_{2}^{(^{1}S_{0})}$, and $L_{1,A}$ are all comparable. The effect is $\sim
10\%$ in the value of $L_{1,A}$ at $\mu =m_{\pi }$ for a natural value of $%
L_{1,A}$ given by dimensional analysis 
\begin{equation}
\left| L_{1,A}\right| \approx {\frac{4\pi }{M}}{\frac{1}{\mu ^{2}}}\approx 5~%
{\rm fm}^{3}\>.  \label{L1a}
\end{equation}
This 10\% uncertainty in $L_{1,A}$ corresponds to a 1\% uncertainty in the
total cross sections. This means that we can treat the symmetry breaking effect on $L_{1,A}$
as higher order, and that we can take $L_{1,A}$ to be the same in all
four channels to the precision of this calculation.

\subsection{Total Cross-Sections}

We are now able to present a systematic and convergent picture of inelastic
neutrino-deuteron scattering in all four channels: neutral-current $\nu _{x}(%
\overline{\nu }_{x})d\rightarrow \nu _{x}(\overline{\nu }_{x})pn$ (NC) with $%
x=e,\mu ,\tau $, and charged-current $\nu _{e}d\rightarrow e^{-}pp$ and $%
\overline{\nu }_{e}d\rightarrow e^{+}nn$ (CC).

We parameterize the cross-sections as 
\[
\sigma (E)=a(E)+b(E)L_{1,A}\>, 
\]
where the coupling constant of the axial two body current $L_{1,A}$ (with $%
\mu =m_{\pi }$) is given in units of fm$^{3}$, and present the results in
tables~\ref{ncresult} and \ref{ccresult}. There are also terms quadratic in $%
L_{1,A}$ at NNLO, but they are not significant for values of $L_{1,A}$
considered here. We will neglect these quadratic terms.

\begin{figure}[t]
\centerline{{\epsfxsize=6.0in \epsfbox{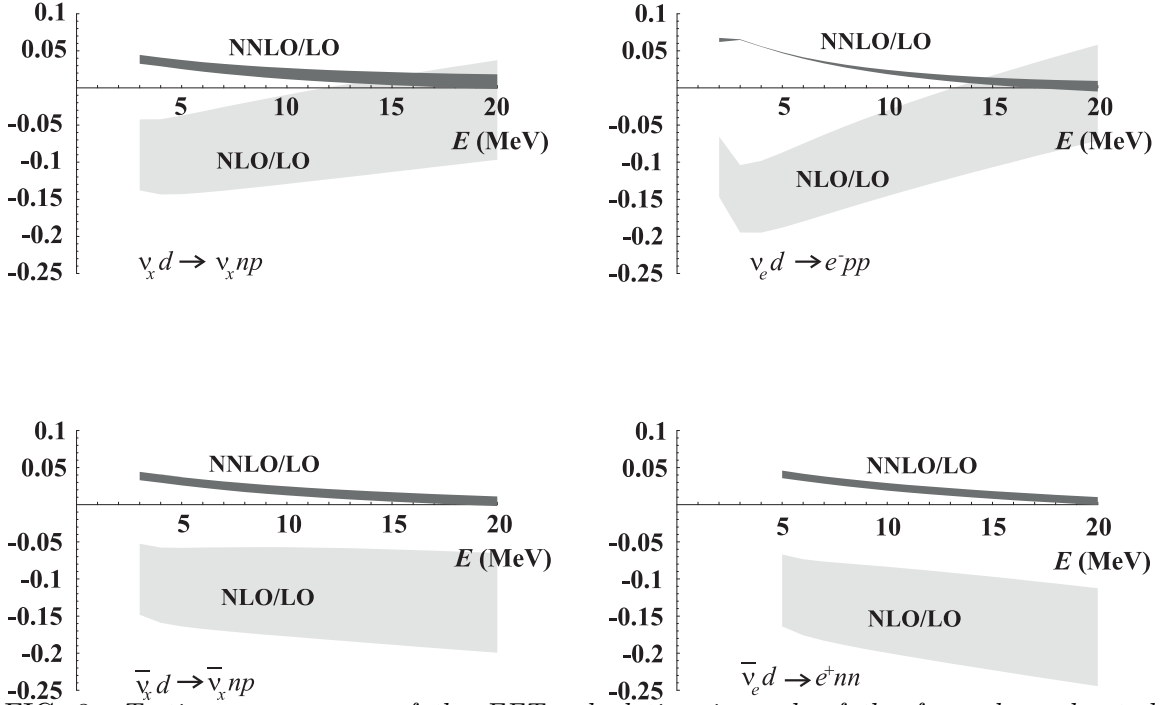}}}
\noindent
\caption{{\it Testing convergence of the EFT calculation in each of the four
channels studied. Shown are the relative NLO and NNLO contributions to the
total cross-section in each channel, for values of $L_{1,A}$ between $-5~%
{\rm fm}^3$ and $+5~{\rm fm}^3$. Note that $\nu_x(\overline\nu_x)$ represents 
$e,\ \mu,\ {\rm or}\tau$
neutrinos(antineutrinos).}}
\label{convergence}
\end{figure}

We have performed this calculation to NNLO largely to test the convergence
of the calculation and, in turn, be able to place constraints on the
theoretical uncertainties in the calculation of $\nu(\overline\nu)$-$d$
breakup. In fig.~\ref{convergence} we compare the size of NLO and NNLO
contributions to the cross-sections against the LO contribution. Given the
uncertainty in $L_{1,A}$, we consider values $-5~{\rm fm}^3\leq L_{1,A}\leq
5~{\rm fm}^3$. The shaded areas represent the range of NLO and NNLO
contributions possible for these values of $L_{1,A}$. We see that the
typical NLO contribution is of order 5-20\%, while the typical NNLO
contribution is less than 5\% and, better still, less than 3\% above 5~MeV.

\begin{figure}[t]
\centerline{{\epsfxsize=6in \epsfbox{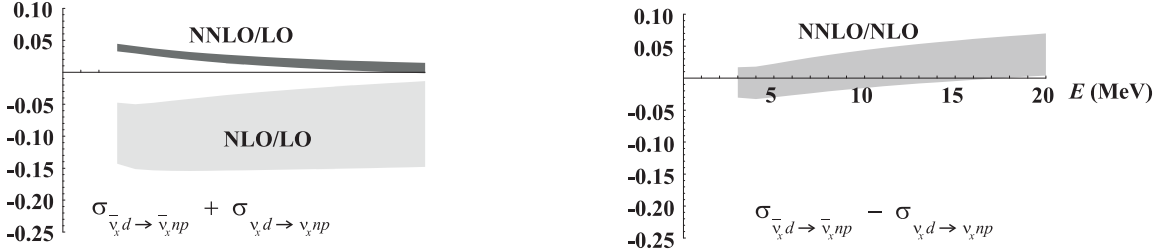}}}
\noindent
\caption{{\it A further test of the convergence of the EFT calculation,
using the NC channel. The left-hand graph arises from comparing
contributions from the sum of the $\protect\nu d$ and $\overline\nu 
d$ channels, or specifically the contributions from $W_1$ and $W_2$ which
received contributions from all three orders of perturbation theory. The
right-hand graph looks at the difference between the two channels, or
specifically the contributions from $W_3$ which appear at NLO and NNLO only.
As in fig~\ref{convergence}, the shaded areas represent the effect of
varying $L_{1,A}$ between $-5~{\rm fm}^3$ and $+5~{\rm fm}^3$. }}
\label{ncconvergence}
\end{figure}

A clearer picture of convergence emerges if we decompose the cross-sections
into a symmetric piece arising from structure factors receiving
contributions at all three orders ($W_1$ and $W_2$), and an antisymmetric
piece receiving contributions only at NLO and NNLO ($W_3$). This is done for
the neutral current cross-sections in fig.~\ref{ncconvergence}, where we see
a cleaner separation between size of the NLO and NNLO contributions. From
this we can expect, with some confidence, that the NNNLO contribution will
be less than 3\%. This in turn represents the formal theoretical uncertainty
in our calculation.

\begin{table}
\begin{tabular}{c|ll|ll}
&\multicolumn{2}{c|}{$\nu_xd\rightarrow \nu_xnp$ ($x=e,\mu,\tau$)}&\multicolumn{2}{c}
{$\overline\nu_xd\rightarrow \overline\nu_xnp$ ($x=e,\mu,\tau$)}\\
$E$ (MeV)&$a$ ($10^{-42}$ cm$^2$)&$b$ ($10^{-42}$ cm$^2$/fm$^3$)&
$a$ ($10^{-42}$ cm$^2$)&$b$ ($10^{-42}$ cm$^2$/fm$^3$)\\
\hline
3 & 0.00315 & 0.000035 & 0.00312 & 0.000036 \cr
4 & 0.0287 & 0.00034 & 0.0282 & 0.00034 \cr
5 & 0.0885 & 0.0011 & 0.0865 & 0.0011 \cr
6 & 0.188 & 0.0024 & 0.182 & 0.0024 \cr  
7 & 0.329 & 0.0044 & 0.318 & 0.0043 \cr
8 & 0.514 & 0.0070 & 0.49 & 0.0069 \cr
9 & 0.744 & 0.010 & 0.710 & 0.010 \cr
10 & 1.02 & 0.015 & 0.968 & 0.014 \cr
11 & 1.34 & 0.019 & 1.27 & 0.019 \cr
12 & 1.72 & 0.025 & 1.61 & 0.025 \cr
13 & 2.13 & 0.032 & 1.99 & 0.031 \cr
14 & 2.60 & 0.039 & 2.41 & 0.038 \cr
15 & 3.12 & 0.047 & 2.87 & 0.046 \cr
16 & 3.69 & 0.056 & 3.37 & 0.054 \cr
17 & 4.31 & 0.066 & 3.91 & 0.064 \cr
18 & 4.97 & 0.077 & 4.49 & 0.074 \cr
19 & 5.69 & 0.089 & 5.11 & 0.085 \cr
20 & 6.47 & 0.102 & 5.76 & 0.097 \cr
\end{tabular}
\caption{{\it Results for the neutral-current cross-sections, calculated
to NNLO.
The total cross-section
for each channel is parameterized as $\sigma(E)=a(E)+b(E)L_{1,A}$.  }}
\label{ncresult}
\end{table}

\begin{table}
\begin{tabular}{c|ll|ll}
&\multicolumn{2}{c|}{$\nu_ed\rightarrow e^-pp$}&
\multicolumn{2}{c}{$\overline\nu_ed\rightarrow e^+nn$}\\
$E$ (MeV)&$a$ ($10^{-42}$ cm$^2$)&$b$ ($10^{-42}$ cm$^2$/fm$^3$)&
$a$ ($10^{-42}$ cm$^2$)&$b$ ($10^{-42}$ cm$^2$/fm$^3$)\\
\hline
2 & 0.00344 & 0.000031 \cr
3 & 0.0439 & 0.00044 \cr
4 & 0.146 & 0.0016 \cr
5 & 0.320 & 0.0036 & 0.0264 & 0.00030 \cr
6 & 0.574 & 0.0067 & 0.110 & 0.0014 \cr
7 & 0.914 & 0.011 & 0.261 & 0.0034 \cr
8 & 1.34 & 0.017 & 0.482 & 0.0065 \cr
9 & 1.87 & 0.024 & 0.776 & 0.011 \cr
10 & 2.48 & 0.032 & 1.14 & 0.016 \cr
11 & 3.20 & 0.042 & 1.58 & 0.023 \cr
12 & 4.01 & 0.054 & 2.09 & 0.032 \cr
13 & 4.93 & 0.067 & 2.68 & 0.041 \cr
14 & 5.95 & 0.082 & 3.33 & 0.052 \cr
15 & 7.08 & 0.098 & 4.06 & 0.065 \cr
16 & 8.31 & 0.12 & 4.85 & 0.079 \cr
17 & 9.66 & 0.14 & 5.71 & 0.094 \cr
18 & 11.12 & 0.16 & 6.63 & 0.11 \cr
19 & 12.70 & 0.18 & 7.63 & 0.13 \cr
20 & 14.39 & 0.21 & 8.68 & 0.15 \cr
\end{tabular}
\caption{{\it Results for the charged-current cross-sections calculated to
NNLO. The total cross-section
for each channel is parameterized as $\sigma(E)=a(E)+b(E)L_{1,A}$. }}

\label{ccresult}
\end{table}

\begin{figure}[!t]
\centerline{{\epsfxsize=3.7 in \epsfbox{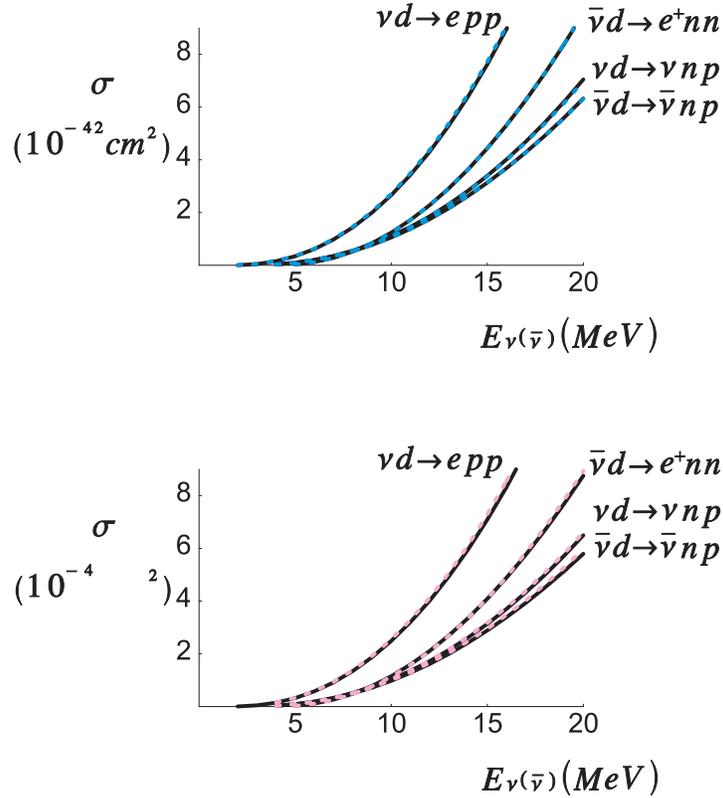}}}
\noindent
\caption{{\it Inelastic $\protect\nu (\overline{\protect\nu })d$
cross-sections as a function of incident $\protect\nu
(\overline{\protect\nu
})$ energy. The solid curves in the upper graph are
NSGK~\protect\cite{NSGK}%
, while the dashed curves are the EFT results at NNLO, fit with
$L_{1,A}=5.6$%
~fm$^{3}$. The solid curves in the lower graph are the results of YHH~
\protect\cite{YHH}, and the dashed curves are the NNLO EFT results fit
with $%
L_{1,A}=0.94$~fm$^{3}$. In both graphs, the dashed curves all lie right on
top of the solid curves.}}
\label{NNLOfit}
\end{figure}

We compare our results to those of the potential model calculations NSGK~ 
\cite{NSGK} and YHH~\cite{YHH} in fig.~\ref{NNLOfit}. We perform a global
fit of our results to these model calculations with $L_{1,A}$ as the only
free parameter. We find that the best fit to NSGK is given by $%
L_{1,A}^{NSGK}=5.6$~fm$^{3}$, and for YHH $L_{1,A}^{YHH}=0.94$~fm$^{3}$.
These values are both consistent with the natural value of $L_{1,A}$
estimated in eq.~(\ref{L1a}) and the quality of the fit is impressive. This
indicates the 5-10\% difference in two potential model calculations is
largely due to different assumptions made about short distance physics.

\begin{figure}[!t]
\centerline{{\epsfxsize=3.5in \epsfbox{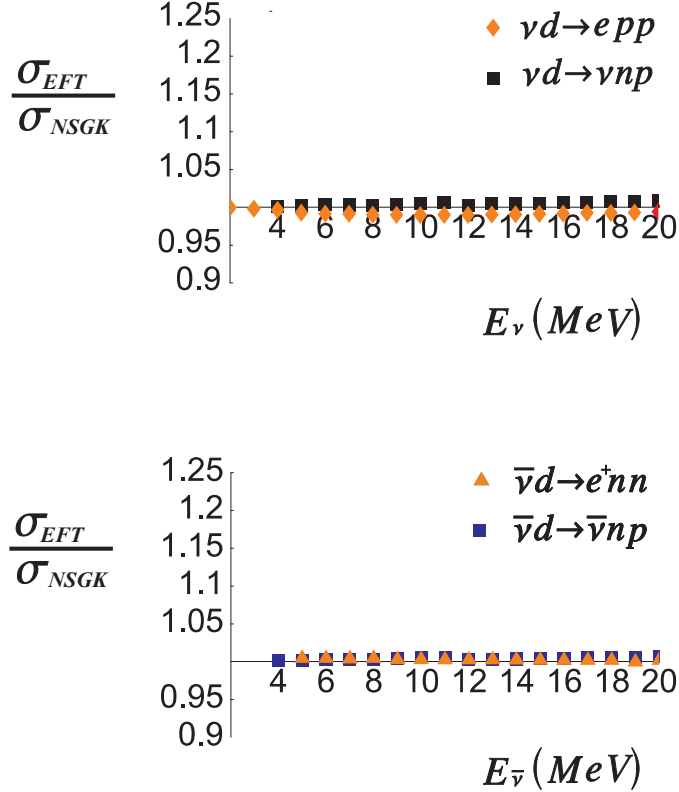}}}
\noindent
\caption{{\it Ratios between the EFT calculation at NNLO and the potential
model result of NSGK~\protect\cite{NSGK}. The upper graph compares the two
channels of $\protect\nu$-$d$ scattering, and the lower graph the two
channels of $\overline\protect\nu$-$d$ scattering. Agreement is better
than
1\% over the whole range of energies shown, for a single value of $%
L_{1,A}=5.6$~fm$^3$. }}
\label{sigratio1}
\end{figure}

\begin{figure}[!t]
\centerline{{\epsfxsize=3.5in \epsfbox{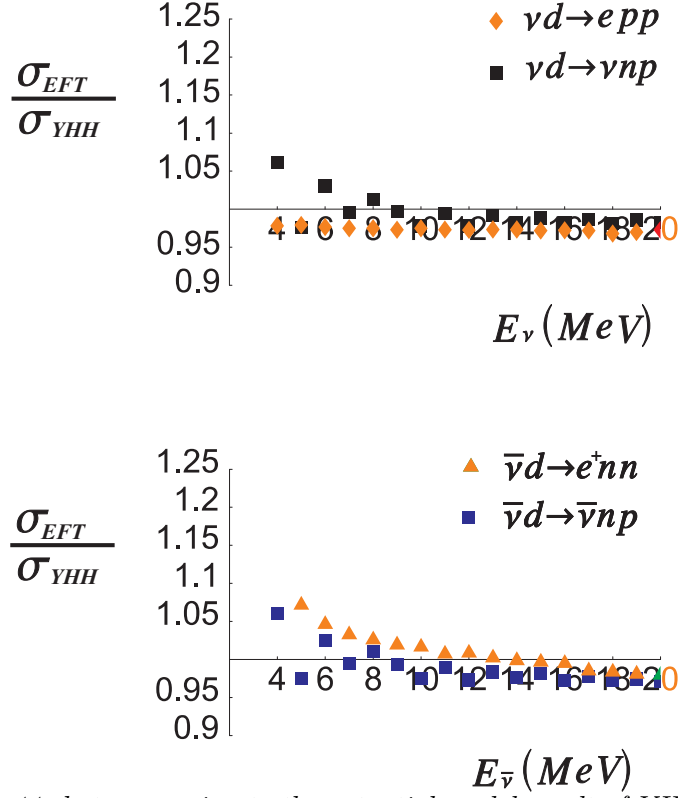}}}
\noindent
\caption{{\it As in fig.~\ref{sigratio1}, but comparing to the potential
model result of YHH~\protect\cite{YHH}. The agreement is not as good here,
using a best-fit value of $L_{1,A}=0.94$~fm$^3$.}}
\label{sigratio2}
\end{figure}

We further examine the ratios of our cross-sections to the potential model
results of NSGK and YHH in figs.~\ref{sigratio1} and \ref{sigratio2}
respectively. We see that there are deviations of order 5\% between our
result and YHH in all four channels. The fluctuations seen are in the
results of YHH, and disagreements at threshold are most likely due to
differences in the effective range parameters that can be associated with
each calculation. However, the agreement between our result at NSGK is quite
impressive - better than 1\% over the whole range of neutrino energies
studied. For comparison, we note the effective range parameters used here
and those calculated from the potential used in NSGK~\cite{sato} (using the
Argonne $v_{18}$ potential~\cite{av18}, but with only Coulomb
electromagnetic interactions) in table~\ref{eretab}.

\begin{table}[tbp]
\begin{tabular}{rccccccc}
& $a^{^1S_0,pp}$ (fm) & $r^{^1S_0,pp}_0$ (fm) & $a^{^1S_0,nn}$ (fm) & $%
r^{^1S_0,nn}_0$ (fm) & $a^{^1S_0,np}$ (fm) & $r^{^1S_0,np}_0$ (fm) & $\rho_d$
(fm) \\ \hline
This Work & -7.82 & 2.79 & -18.5 & 2.80 & -23.7 & 2.73 & 1.764 \\ 
NSGK & -7.815 & 2.78 & -18.5 & 2.83 & -23.73 & 2.69 & 1.767
\end{tabular}
\caption{{\it Effective range parameters as used in our work and NSGK~ 
\protect\cite{NSGK}.}}
\label{eretab}
\end{table}

\begin{figure}[!t]
\centerline{{\epsfxsize=3.5in \epsfbox{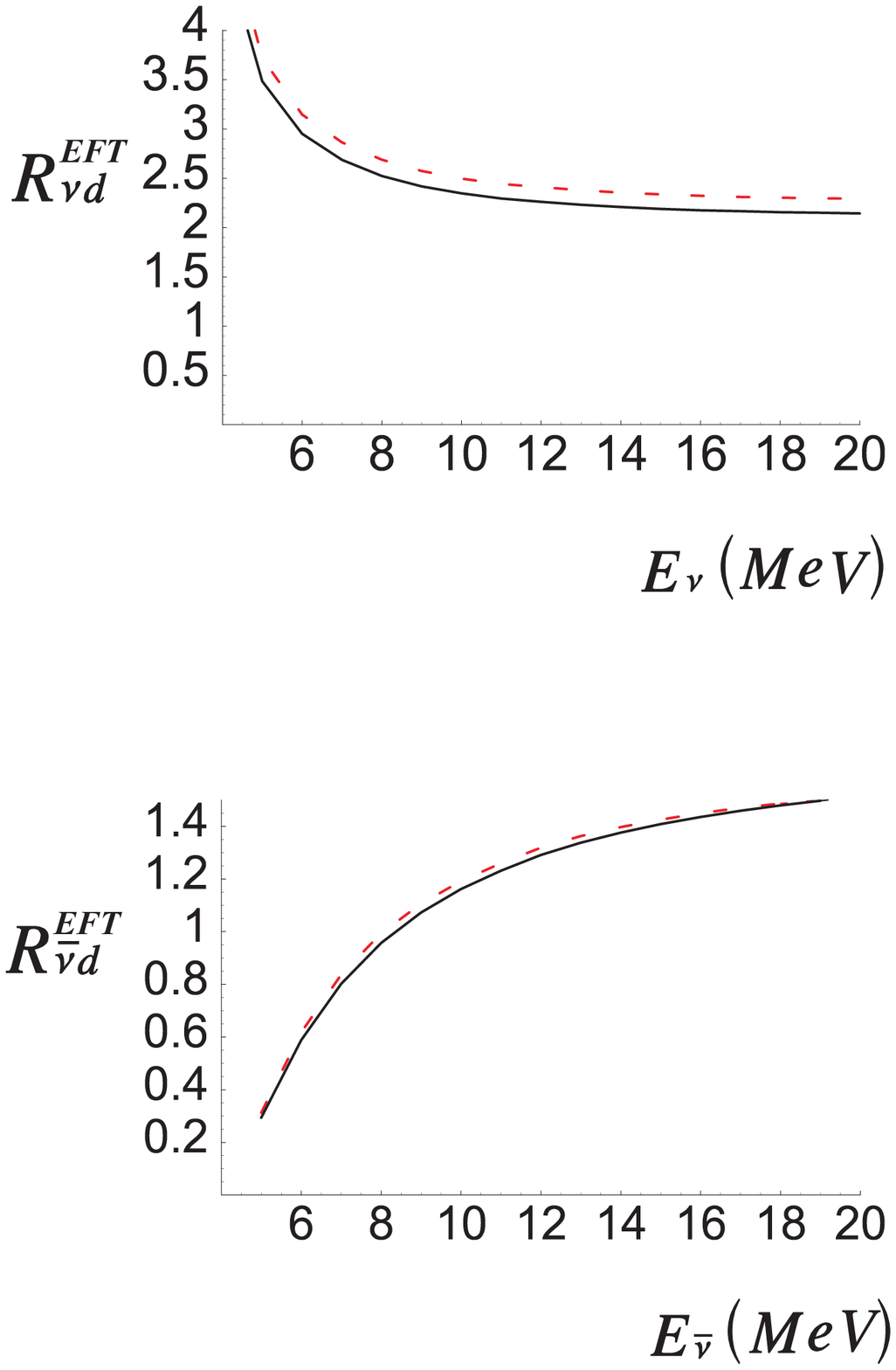}}}
\noindent
\caption{{\it $CC$ to $NC$ cross-section ratios for $\protect\nu -d$ and $%
\overline{\protect\nu }-d$ scattering, as functions of incident energy.
Shown are ratios of the EFT NNLO results with $L_{1,A}=-20$~fm$^{3}$
(dashed) and $L_{1,A}=40$~fm$^{3}$, demonstrating the insensitivity of the
ratio to the value of $L_{1,A}$.}}
\label{CCNCratio1}
\end{figure}

\begin{figure}[!t]
\centerline{{\epsfxsize=3.75 in \epsfbox{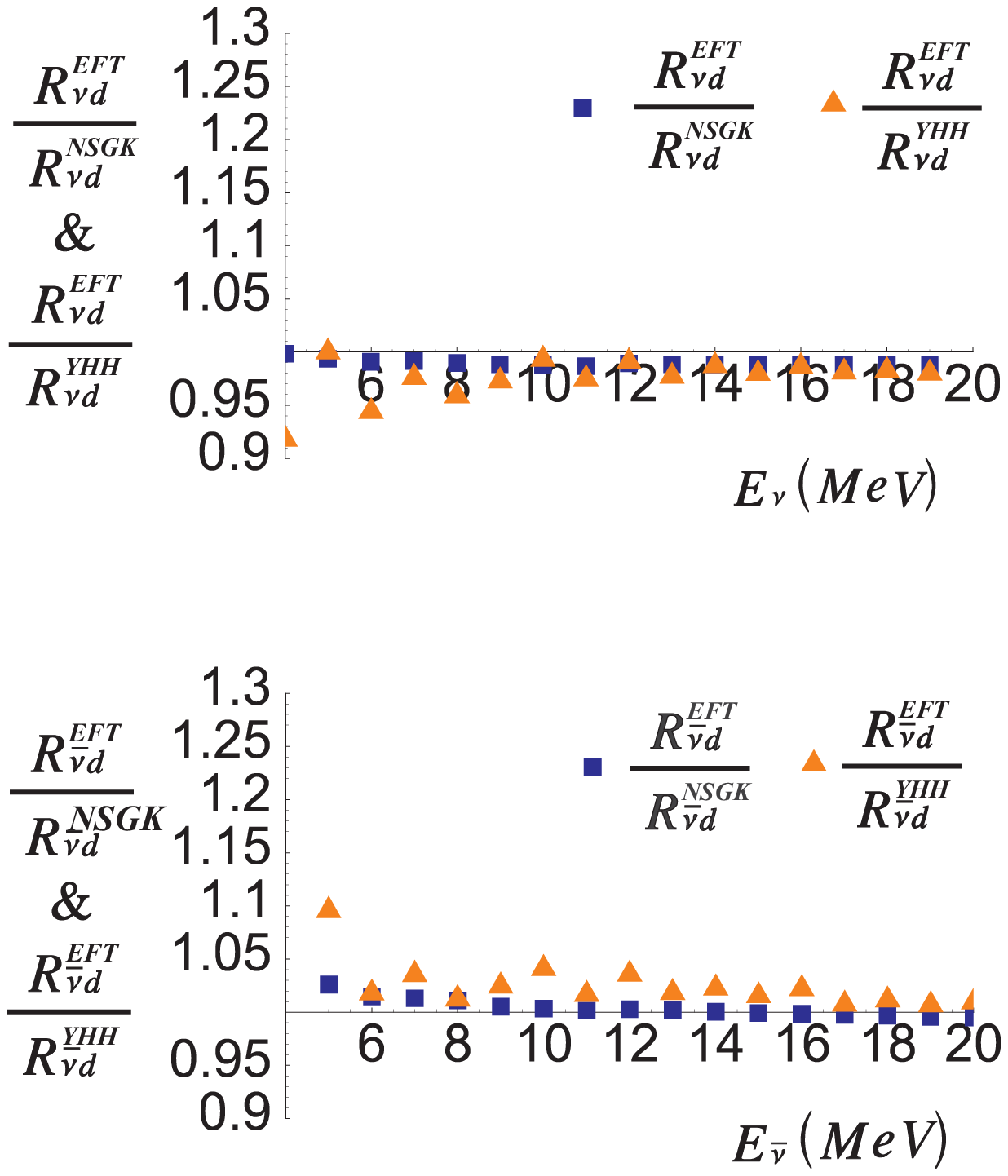}}}
\noindent
\caption{{\it The `ratio of ratios' between the EFT results at NNLO and
those of NSGK~\protect\cite{NSGK} and YHH~\protect\cite{YHH}, with the EFT
results using a median value of $L_{1,A}=3.7$~fm$^{3}$. The upper graph
compares the ratios for $\protect\nu -d$ scattering, and the lower graph the
ratios for $\overline{\protect\nu }-d$ scattering. Again, the agreement with
NSGK is excellent. }}
\label{CCNCratio2}
\end{figure}

Of import to SNO is the ratio between charged and neutral current
cross-sections 
\begin{equation}
R\equiv {\frac{\sigma _{CC}}{\sigma _{NC}}}\>.
\end{equation}
As shown in ref.~\cite{BC1}, this ratio $R$ at NLO was insensitive to the
value of $L_{1,A}$, and that is still true at NNLO as seen in fig.~\ref
{CCNCratio1}. We consider two ratios, for both the $\nu $-$d$ (relevant to
SNO) and $\overline{\nu }$-$d$ channels. The latter was the only channel
discussed in ref.~ \cite{BC1}. Variations of $L_{1,A}$ over a large range,
from $-20$ to $+40$~fm$^{3}$, leads to a 6\% variation in $R$. This likely
represents an extreme variation, as it leads to as much as a 90\% change in
the total cross sections. The actual uncertainty in $R$ is almost certainly much
less. Further, we can see that our values of $R$ agree well with the
potential model results (fig.~\ref{CCNCratio2}), with the worst agreement
being seen when comparing to YHH. We use a median value of $L_{1,A}=3.7$~fm$%
^{3}$ for this comparison. The poor agreement with YHH is biased towards
threshold, and the likely reasons for this have already been discussed.


\section{Summary and Conclusions}

We have performed a calculation to NNLO of all four channels of $\nu d$ and $%
\overline\nu d$ breakup. Our work agrees very well with the latest potential
model calculations, subject to the fitting of a single unknown counterterm $%
L_{1,A}$. Working at NNLO has allowed us to determine that our calculation
indeed converges at the neutrino energies of interest, which in turn allows
us to determine a formal theoretical uncertainty of 3\% in our calculation.
The only outstanding issue continues to be the determination of $L_{1,A}$.

It is imperative that an experimental determination of this counterterm be
made. The theory without pions cannot reach energies that would allow this
to be done with SAMPLE, but a breakthrough in the treatment of pions at NNLO
might make access possible in the full theory. For now, ORLAND offers the
best hope, with plans for a high-precision measurement in the $\nu $-$d$ CC
channel.

\vskip2in \centerline{\bf ACKNOWLEDGMENTS} We would like to thank Martin
Savage, Tom Cohen, John Beacom and Hamish Robertson for useful discussions. We would like to
\ thank K.\ Kubodera for providing us with the results of their new
potential model calculation (NSGK) and T.\ Sato for providing information on
that calculation's effective range parameters. We thank the Department of
Physics at the University of Washington where this work was initiated, the Institute for 
Nuclear Theory at the University of Washington for
its hospitality and the Department of Energy for partial support during the
completion of this work. J.-W.C.\ is supported, in part, by the Department
of Energy under grant DOE/ER/40762-213. M.B.\ and X.K.\ are supported by grants from
the Natural Sciences and Engineering Research Council of Canada.

\appendix

\section{Fitting the Parameters of Effective Field Theory}
\label{appa}

Here, we summarize the fits to parameters in the theory without pions.

\subsection{The $^3S_1$ channel}

In the $^{3}S_{1}$ channel, we use the effective range expansion about the
deuteron pole. Thus, 
\begin{equation}
p\cot \delta _{0}=-\gamma +{\frac{1}{2}}\rho _{d}(p^{2}+\gamma ^{2})+\ ...
\end{equation}
To keep the deuteron pole position unchanged at each order in perturbation
theory, we expand 
\begin{eqnarray}
C_{0}^{(\,^{3}S_{1})}
&=&C_{0,-1}^{(\,^{3}S_{1})}+C_{0,0}^{(\,^{3}S_{1})}+C_{0,1}^{(\,^{3}S_{1})}+%
\ ...  \nonumber \\
C_{2}^{(\,^{3}S_{1})}
&=&C_{2,-2}^{(\,^{3}S_{1})}+C_{2,-1}^{(\,^{3}S_{1})}+...  \nonumber \\
C_{4}^{(\,^{3}S_{1})} &=&C_{4,-3}^{(\,^{3}S_{1})}+\ ...
\end{eqnarray}
where the first index continues to denote the momentum dependence while the
second index indicates the explicit power counting in the $Q$-expansion.
Using the definition of the scattering amplitude 
\begin{equation}
{\cal A}={%
{\displaystyle{4\pi  \over M_{N}}}%
}{\frac{1}{p\cot \delta _{0}-ip}}
\end{equation}
along with the amplitude computed using the power divergent subtraction
(PDS) scheme proposed by KSW~\cite{KSW}
\begin{equation}
{\cal A}_{EFE}=-{\frac{1}{{%
{\displaystyle{1 \over C}}%
}+{%
{\displaystyle{M_{N} \over 4\pi }}%
}(\mu +ip)}}
\end{equation}
with $C=\sum C_{2n}^{(\,^{3}S_{1})}p^{2n}$.

Matching terms order-by-order in a $Q$-expansion, one obtains
\begin{eqnarray}
C_{0,-1}^{(\,^{3}S_{1})} &=&{\frac{-4\pi }{M_{N}}}{\frac{1}{(\mu -\gamma )}}
\nonumber \\
C_{0,0}^{(\,^{3}S_{1})} &=&{\frac{2\pi }{M_{N}}}{\frac{\rho _{d}\gamma ^{2}}{%
(\mu -\gamma )^{2}}}  \nonumber \\
C_{0,1}^{(\,^{3}S_{1})} &=&-{\frac{\pi }{M_{N}}}{\frac{\rho _{d}^{2}\gamma
^{4}}{(\mu -\gamma )^{3}}}  \nonumber \\
C_{2,-2}^{(\,^{3}S_{1})} &=&{\frac{2\pi }{M_{N}}}{\frac{\rho _{d}}{(\mu
-\gamma )^{2}}}  \nonumber \\
C_{2,-1}^{(\,^{3}S_{1})} &=&-{\frac{2\pi }{M_{N}}}{\frac{\rho _{d}^{2}\gamma
^{2}}{(\mu -\gamma )^{3}}}  \nonumber \\
C_{4,-3}^{(\,^{3}S_{1})} &=&-{\frac{\pi }{M_{N}}}{\frac{\rho _{d}^{2}}{(\mu
-\gamma )^{3}}}\>,
\end{eqnarray}
where we have neglected relativistic corrections.

\subsection{The $^1S_0$ channel}

We will deal with the $pp$ channel separately in the next section. For the $%
np$ and $nn$ channels, the procedure is somewhat simpler. The effective
range expansion is given by 
\begin{equation}
p\cot \delta _{0}=-{\frac{1}{a}}+{\frac{1}{2}}r_{0}p^{2}+\
...
\end{equation}
Here, one obtains
\begin{eqnarray}
C_{0}^{(\,^{1}S_{0})} &=&{\frac{-4\pi }{M_{N}}}{\frac{1}{\bigg(\mu -{%
{\displaystyle{1 \over a^{(\,^{1}S_{0})}}}%
}\bigg)}}  \nonumber \\
C_{2}^{(\,^{1}S_{0})} &=&{\frac{2\pi }{M_{N}}}{\frac{r_{0}^{(\,^{1}S_{0})}}{%
\bigg(\mu -{%
{\displaystyle{1 \over a^{(\,^{1}S_{0})}}}%
}\bigg)^{2}}}  \nonumber \\
C_{4}^{(\,^{1}S_{0})} &=&-{\frac{\pi }{M_{N}}}{\frac{r_{0}^{(%
\,^{1}S_{0})^{2}}}{\bigg(\mu -{%
{\displaystyle{1 \over a^{(\,^{1}S_{0})}}}%
}\bigg)^{3}}}\>.
\end{eqnarray}

\subsection{The $^1S_0$ $pp$ channel}

In this section we show how we fix the $pp$ scattering parameters to phase
shift data through matching on to effective range expansion.

The S-wave $pp$ scattering amplitude can be decomposed into 
\begin{equation}
{\cal A}={\cal A}_{C}+{\cal A}_{SC}\qquad ,
\end{equation}
where ${\cal A}_{C}$ is the pure Coulomb interaction amplitude with
strong interaction ``turned off'' and ${\cal A}_{SC}$ is the remaining part
with both strong and Coulomb interactions. Phase shifts $\delta $ and $%
\delta _{C}$ are defined by 
\begin{eqnarray}
{\cal A} &=&\frac{4\pi }{M_{N}}\frac{e^{2i\delta }-1}{2ip}\qquad ,  \nonumber
\\
{\cal A}_{C} &=&\frac{4\pi }{M_{N}}\frac{e^{2i\delta _{C}}-1}{2ip}\qquad .
\end{eqnarray}
Then 
\begin{equation}
{\cal A}_{SC}=\frac{4\pi }{M_{N}}e^{2i\delta _{C}}\left( \frac{e^{2i\delta
_{SC}}-1}{2ip}\right) =\frac{4\pi }{M_{N}}\frac{e^{2i\delta _{C}}}{p\left(
\cot \delta _{SC}-i\right) }\qquad ,  \label{match1}
\end{equation}
where $\delta _{SC}\equiv \delta -\delta _{C}.$ The Effective Range
Expansion states that 
\begin{equation}
p\left( \cot \delta _{SC}-i\right) =\left( -\frac{1}{a^{(pp)}}+\frac{1}{2}%
r_{0}^{(pp)}p^{2}+v^{(pp)}p^{4}+\cdots \right) -\alpha M_{N}H(\eta )\qquad ,
\label{match2}
\end{equation}
where $v^{(pp)}$ is the shape parameter. This expansion is related to ${\cal %
A}_{SC}/e^{2i\delta _{C}}$ which can be thought of as the $pp$ scattering
amplitude with the pure Coulomb phase shift removed.
The sum of the diagrams give 
\begin{equation}
\frac{{\cal A}_{SC}}{e^{2i\delta _{C}}}\equiv A^{(1S_{0},pp)}=-%
{\displaystyle{1 \over %
{\displaystyle{1 \over C}}-J_{0}}}%
\qquad ,  \label{match3}
\end{equation}
where 
\begin{equation}
C=\sum\limits_{n=0}^{\infty }C_{2n}^{(pp)}(p^{2}-\alpha M_{N}\mu )^{n}\qquad
,
\end{equation}
and 
\begin{equation}
J_{0}=\frac{\alpha M_{N}^{2}}{4\pi }\left( \ln \frac{\mu \sqrt{\pi }}{\alpha
M_{N}}+1-\frac{3}{2}\gamma _{E}\right) -\frac{\mu M_{N}}{4\pi }-\frac{\alpha
M_{N}^{2}}{4\pi }H(\eta )\qquad  \label{J0}
\end{equation}
with the Euler's constant $\gamma _{E}=0.577$. In eq.\thinspace (\ref{J0}),
a 4-dimensional pole $\alpha M_{N}^{2}/(4\pi (4-d))$, along with the
3-dimensional pole, are subtracted from $J_{0}$ using the
PDS prescription. 

It is convenient to insert the expansion parameter $\varepsilon $ into
eqs.\thinspace (\ref{match1}-\ref{match3}) to keep track of the $Q$
expansion. Then the effective field theory parameters $C_{2n}^{(pp)}$ can be
solved by matched on to the effective range expansion order-by-order in $%
\varepsilon $. The insertion of $\varepsilon $ is done by the transformation 
\begin{equation}
\begin{array}{cc}
{\displaystyle{1 \over a^{(pp)}}}%
\rightarrow 
{\displaystyle{1 \over a^{(pp)}}}%
\varepsilon \qquad & p\rightarrow p\varepsilon \qquad \\ 
r_{0}^{(pp)}\rightarrow r_{0}^{(pp)}\qquad & v^{(pp)}\rightarrow v^{(pp)}\
\qquad \\ 
\alpha M_{N}\rightarrow \alpha M_{N}\varepsilon ^{2}\qquad & \mu \rightarrow
\mu \varepsilon \qquad \\ 
H(\eta )\rightarrow H(\eta )%
{\displaystyle{1 \over \varepsilon }}%
\qquad & J_{0}\rightarrow J_{0}\varepsilon \qquad
\end{array}
\label{expand1}
\end{equation}
The assignment of powers in $\varepsilon $ reflects the powers in $Q$
scaling of those parameters. For example, $1/a^{(pp)}$ and $p$ scale like $Q$%
, $\alpha M_{N}$ scales like $Q^{2}$ while $\left| H(\eta )\right| \leq
p/(\alpha M_{N})$ scales like $1/\varepsilon $. The $C_{2n}^{(pp)}$ can be
represented in a manner analogous that seen in the $^{3}S_{1}$ channel~\cite
{CRS} 
\begin{eqnarray}
C_{0}^{(pp)} &=&C_{0,-1}^{(pp)}\varepsilon
^{-1}+C_{0,0}^{(pp)}+C_{0,1}^{(pp)}\varepsilon +\cdots  \nonumber \\
C_{2}^{(pp)} &=&C_{2,-2}^{(pp)}\varepsilon ^{-2}+C_{2,-1}^{(pp)}\varepsilon
^{-1}\ +\cdots  \nonumber \\
C_{4}^{(pp)} &=&C_{4,-3}^{(pp)}\varepsilon ^{-3}+\cdots \qquad .
\label{expand2}
\end{eqnarray}
Then from \thinspace (\ref{match1}-\ref{match3}) we obtain 
\begin{equation}
C_{0,-1}^{(pp)}=\frac{4\pi }{M_{N}}\frac{1}{%
{\displaystyle{1 \over a^{(pp)}}}%
-\mu +\alpha M_{N}\left( \ln 
{\displaystyle{\mu \sqrt{\pi } \over \alpha M_{N}}}%
+1-\frac{3}{2}\gamma _{E}\right) }
\end{equation}
and 
\begin{equation}
\begin{array}[t]{ll}
C_{0,0}^{(pp)}=0\qquad \qquad & C_{0,1}^{(pp)}=\alpha M_{N}\mu
C_{2,-2}^{(pp)}\qquad \\ 
C_{2,-2}^{(pp)}=%
{\displaystyle{M_{N} \over 8\pi }}%
r_{0}^{(pp)}C_{0,-1}^{(pp)^{2}}\qquad & C_{2,-1}^{(pp)}=0\qquad \\ 
C_{4,-3}^{(pp)}=%
{\displaystyle{M_{N}^{2} \over 64\pi ^{2}}}%
r_{0}^{(pp)^{2}}C_{0,-1}^{(pp)^{3}}\qquad & 
\end{array}
\end{equation}
The solutions of parameters for $nn$ and $np^{(^{1}S_{0})}$ given in the
previous section can be obtained by taking $\alpha =0$ from the above
expressions.

\bigskip 

\section{Summary of Scattering Cross Section}
\label{appb}

In this appendix, we explicitly list the relevant formulae for the four $\nu
(\overline{\nu })d$ breakup processes. To make the formulae compact, certain
higher order terms have been resummed. The $Q$-expansion expressions to NNLO
shown in the text can always be recovered by expanding in $\epsilon $ to ${\cal %
O}(\epsilon ^{2})$. Here $\epsilon $ is just a device to keep track of the
$Q$-expansion -- its value should be set to $1$ after the expansion is
performed.

\subsection{$\protect\nu +d\rightarrow \protect\nu +n+p$ and $\overline{%
\protect\nu }+d\rightarrow \protect\nu +n+p$}

\noindent
Differential cross section: 
\begin{equation}
{\displaystyle{d^{2}\sigma  \over d\omega ^{^{\prime }}d\Omega }}%
=%
{\displaystyle{G_{F}^{2}\omega ^{\prime }\left| {\bf k}^{\prime }\right|  \over 2\pi ^{2}}}%
\left[ 2W_{1}\sin ^{2}%
{\displaystyle{\theta  \over 2}}%
+W_{2}\cos ^{2}%
{\displaystyle{\theta  \over 2}}%
\mp 2%
{\displaystyle{(\omega +\omega ^{\prime }) \over M_{d}}}%
W_{3}\sin ^{2}%
{\displaystyle{\theta  \over 2}}%
\right] \quad ,
\end{equation}
with $-(+)$ sign for the $\nu (\overline{\nu })d$ scattering.
\medskip

\noindent
Phase space: 
\begin{gather}
\text{Max}\left[ -1,1-%
{\displaystyle{4M_{N}(\nu -B)-\nu ^{2} \over 2\omega \omega ^{\prime }}}%
\right] \leq \cos \theta \leq 1\quad ,  \nonumber \\
0\leq \omega ^{\prime }\leq \omega -2(M_{N}-\sqrt{M_{N}^{2}-\gamma ^{2}}%
)\quad .
\end{gather}
Structure functions (as mentioned above, one can obtain the NNLO results by
expanding $\epsilon $ to ${\cal O}(\epsilon ^{2})$ ): 
\begin{eqnarray}
W_{1} &=&\frac{1}{\left( 1-\epsilon \gamma \rho _{d}\right) }\left[
2(C_{A}^{(0)^{2}}+C_{A}^{(1)^{2}})F_{1}+%
{\displaystyle{1 \over 3}}%
(C_{A}^{(0)^{2}}-C_{A}^{(1)^{2}})F_{2}+\frac{8}{3}C_{A}^{(0)^{2}}F_{3}+\frac{%
4}{3}C_{A}^{(1)^{2}}F_{4}+G_{2}\right] \quad ,  \nonumber \\
W_{2} &=&W_{1}+\frac{1}{\left( 1-\epsilon \gamma \rho _{d}\right) }\left[
2(C_{V}^{(0)^{2}}+C_{V}^{(1)^{2}})F_{1}+(C_{V}^{(0)^{2}}-C_{V}^{(1)^{2}})F_{2}+4C_{V}^{(0)^{2}}F_{3}+G_{1}%
\right] \quad ,  \nonumber \\
W_{3} &=&\frac{-2\epsilon }{\left( 1-\epsilon \gamma \rho _{d}\right) }\left[
2(C_{A}^{(0)}C_{M}^{(0)}+C_{A}^{(1)}C_{M}^{(1)})F_{1}+%
{\displaystyle{1 \over 3}}%
(C_{A}^{(0)}C_{M}^{(0)}-C_{A}^{(1)}C_{M}^{(1)})F_{2}+\frac{8}{3}%
C_{A}^{(0)}C_{M}^{(0)}F_{3}\right.  \nonumber \\
&&\left. +\frac{4}{3}C_{A}^{(1)}C_{M}^{(1)}F_{4}+G_{3}\right]
\end{eqnarray}
\begin{equation}
\begin{array}[t]{ll}
C_{V}^{(0)}=-\sin ^{2}\vartheta _{W} & ,\quad C_{V}^{(1)}=%
{\displaystyle{1 \over 2}}%
(1-2\sin ^{2}\vartheta _{W})\quad , \\ 
C_{A}^{(0)}=-%
{\displaystyle{1 \over 2}}%
\Delta s & ,\quad C_{A}^{(1)}=%
{\displaystyle{1 \over 2}}%
g_{A}\quad , \\ 
C_{M}^{(0)}=-2\sin ^{2}\vartheta _{W}\kappa ^{(0)}-\frac{1}{2}\mu _{S} & 
,\quad C_{M}^{(1)}=(1-2\sin ^{2}\vartheta _{W})\kappa ^{(1)}\quad .
\end{array}
\end{equation}
\begin{eqnarray}
F_{1} &=&%
{\displaystyle{2M_{N}\gamma \ p \over \pi \left( p^{2}+\gamma ^{2}\right) ^{2}}}%
\left( 1+\frac{{\bf q}^{2}\left( p^{2}-\gamma ^{2}\right) }{2\left(
p^{2}+\gamma ^{2}\right) ^{2}}\right) \quad ,\quad F_{2}=%
{\displaystyle{4M_{N}\gamma \ p \over \pi \left( p^{2}+\gamma ^{2}\right) ^{2}}}%
\left( 1-\frac{{\bf q}^{2}\left( p^{2}+3\gamma ^{2}\right) }{6\left(
p^{2}+\gamma ^{2}\right) ^{2}}\right) \quad ,  \nonumber \\
F_{3} &=&%
{\displaystyle{1\  \over \pi }}%
\text{Im}\left[ B_{0}{}^{2}\left( A_{-1}^{(^{3}S_{1})}(p)+\epsilon
A_{0}^{(^{3}S_{1})}(p)+\epsilon ^{2}A_{1}^{(^{3}S_{1})}(p)\right) \right]
\quad ,  \nonumber \\
F_{4} &=&F_{3}(^{3}S_{1}\rightarrow \ ^{1}S_{0},np)\ \quad .
\end{eqnarray}
\begin{equation}
B_{0}={\bf -}\sqrt{\frac{\gamma }{2\pi }}\frac{M_{N}}{\gamma -ip}\left( 1-%
\frac{{\bf q}^{2}}{12\left( \gamma -ip\right) ^{2}}\right)
\end{equation}
\begin{equation}
p=\sqrt{M_{N}\nu -\gamma ^{2}-\frac{{\bf q}^{2}}{4}+i\epsilon }\quad ,
\end{equation}

\begin{eqnarray}
G_{1} &=&\epsilon C_{V}^{(0)^{2}}%
{\displaystyle{2M_{N}\rho _{d} \over \pi }}%
\mathop{\rm Im}%
\left[ \sqrt{\frac{2\gamma }{\pi }}B_{0}\left(
A_{-1}^{(^{3}S_{1})}(p)+\epsilon A_{0}^{(^{3}S_{1})}(p)\right) \right] \quad
,  \nonumber \\
G_{2} &=&-\epsilon 
{\displaystyle{M_{N} \over 3\pi ^{2}}}%
\mathop{\rm Im}%
\left[ \sqrt{\frac{\gamma }{2\pi }}B_{0}\left[ C_{A}^{(1)}\widetilde{L}%
_{1,A}\left( A_{-1}^{(^{1}S_{0},np)}(p)+\epsilon
A_{0}^{(^{1}S_{0},np)}(p)\right) +4C_{A}^{(0)}\widetilde{L}_{2,A}\left(
A_{-1}^{(^{3}S_{1})}(p)+\epsilon A_{0}^{(^{3}S_{1})}(p)\right) \right] %
\right]  \nonumber \\
&&+\epsilon ^{2}%
{\displaystyle{M_{N}^{2}\gamma  \over 96\pi ^{4}}}%
\mathop{\rm Im}%
\left[ \left( \widetilde{L}_{1,A}^{2}A_{-1}^{(^{1}S_{0},np)}(p)+8\widetilde{L%
}_{2,A}^{2}A_{-1}^{(^{3}S_{1})}(p)\right) \right] \quad , \\
G_{3} &=&\epsilon ^{2}%
{\displaystyle{M_{N} \over 3\pi ^{2}}}%
\mathop{\rm Im}%
\left\{ \sqrt{\frac{\gamma }{2\pi }}B_{0}(p,\left| {\bf q}\right| {\bf )}%
\left[ \left( C_{M}^{(1)}\widetilde{L}_{1,A}+C_{A}^{(1)}\widetilde{L}_{1,M}\
\right) A_{-1}^{(^{1}S_{0},np)}(p)\right. \right.  \nonumber \\
&&\left. \left. +4\left( C_{M}^{(0)}\widetilde{L}_{2,A}+C_{A}^{(0)}%
\widetilde{L}_{2,M}\right) A_{-1}^{(^{3}S_{1})}(p)\right] \right\}  \nonumber
\end{eqnarray}
\begin{eqnarray}
\widetilde{L}_{1,A} &=&-\frac{4\pi \left( \mu -\gamma \right) }{%
M_{N}C_{0}^{(^{1}S_{0},np)}}\left[ L_{1,A}-2\pi C_{A}^{(1)}\left( \frac{M_{N}%
}{2\pi }C_{2}^{(^{1}S_{0},np)}+\frac{\rho _{d}}{\left( \mu -\gamma \right)
^{2}}\right) \right] \quad ,  \nonumber \\
\widetilde{L}_{2,A} &=&\left( \mu -\gamma \right) ^{2}L_{2,A}-2\pi
C_{A}^{(0)}\rho _{d}\quad .
\end{eqnarray}
\begin{eqnarray}
\widetilde{L}_{1,M} &=&-\frac{8\pi \left( \mu -\gamma \right) }{%
M_{N}C_{0}^{(^{1}S_{0},np)}}\left[ (1-2\sin ^{2}\vartheta
_{W})M_{N}L_{1}-\pi C_{M}^{(1)}\left( \frac{M_{N}}{2\pi }%
C_{2}^{(^{1}S_{0},np)}+\frac{\rho _{d}}{\left( \mu -\gamma \right) ^{2}}%
\right) \right] \quad ,  \nonumber \\
\widetilde{L}_{2,M} &=&\ -4\sin ^{2}\vartheta _{W}\left( \mu -\gamma \right)
^{2}M_{N}L_{2}-2\pi C_{M}^{(0)}\rho _{d}\quad ,
\end{eqnarray}
evaluated at $\mu=m_\pi$, with $C_0^{(\,^1S_0,np)}=-3.56$~fm$^2$ and 
$C_2^{(\,^1S_0,np)}=6.55$~fm$^4$. Further, $\Delta s=-0.17$, and
$\mu_s=L_{2,A}=0$ are used in our calculation (though the results are
not sensitive to these choices).

\subsection{$\overline{\protect\nu }+d\rightarrow e^{+}+n+n$}

\noindent
Differential cross section: 
\begin{equation}
{\displaystyle{d^{2}\sigma  \over d\omega ^{^{\prime }}d\Omega }}%
=%
{\displaystyle{G_{F}^{2}\omega ^{\prime }\left| {\bf k}^{\prime }\right|  \over 2\pi ^{2}}}%
\left[ 2W_{1}\sin ^{2}%
{\displaystyle{\theta  \over 2}}%
+W_{2}\cos ^{2}%
{\displaystyle{\theta  \over 2}}%
+2%
{\displaystyle{(\omega +\omega ^{\prime }) \over M_{d}}}%
W_{3}\sin ^{2}%
{\displaystyle{\theta  \over 2}}%
\right] \quad ,
\end{equation}

\noindent
Phase space: 
\begin{gather}
\text{Max}\left[ -1,-{\displaystyle{\frac{4M_{N}(\nu -B+m_{n}-m_{p})-\omega
^{2}-\omega ^{\prime 2}+m_{e}^{2}}{2\omega \sqrt{\omega ^{\prime 2}-m_{e}^{2}%
}\ }}}\right] \leq \cos \theta \leq 1\quad ,  \nonumber \\
m_{e}\leq \omega ^{\prime }\leq \omega -2(M_{N}-\sqrt{%
M_{N}^{2}-M_{N}(B-m_{n}+m_{p})})
\end{gather}

\noindent
Structure functions: 
\begin{eqnarray}
W_{1} &=&\frac{1}{\left( 1-\epsilon \gamma \rho _{d}\right) }\left[
2C_{A}^{(1)^{2}}F_{1}-%
{\displaystyle{1 \over 3}}%
C_{A}^{(1)^{2}}F_{2}+\frac{4}{3}C_{A}^{(1)^{2}}F_{4}+G_{2}\right] \quad , 
\nonumber \\
W_{2} &=&W_{1}+\frac{1}{\left( 1-\epsilon \gamma \rho _{d}\right) }\left[
2C_{V}^{(1)^{2}}F_{1}-C_{V}^{(1)^{2}}F_{2}\right] \quad ,  \nonumber \\
W_{3} &=&\frac{-2\epsilon }{\left( 1-\epsilon \gamma \rho _{d}\right) }\left[
2C_{A}^{(1)}C_{M}^{(1)}F_{1}-%
{\displaystyle{1 \over 3}}%
C_{A}^{(1)}C_{M}^{(1)}F_{2}+\frac{4}{3}C_{A}^{(1)}C_{M}^{(1)}F_{4}+G_{3}%
\right] \quad .
\end{eqnarray}
\begin{equation}
p=\sqrt{M_{N}\left( \nu +m_{n}-m_{p}\right) -\gamma ^{2}-\frac{{\bf q}^{2}}{4%
}+i\epsilon }\quad ,
\end{equation}
\begin{equation}
C_{V}^{(1)}=%
{\displaystyle{\left| V_{ud}\right|  \over \sqrt{2}}}%
\quad ,\quad C_{A}^{(1)}=%
{\displaystyle{\left| V_{ud}\right|  \over \sqrt{2}}}%
g_{A}\quad ,\quad C_{M}^{(1)}=\sqrt{2}\left| V_{ud}\right| \kappa ^{(1)}\quad
\end{equation}
\begin{eqnarray}
F_{1} &=&%
{\displaystyle{2M_{N}\gamma \ p \over \pi \left( p^{2}+\gamma ^{2}\right) ^{2}}}%
\left( 1+\frac{{\bf q}^{2}\left( p^{2}-\gamma ^{2}\right) }{2\left(
p^{2}+\gamma ^{2}\right) ^{2}}\right) \quad ,\quad F_{2}=%
{\displaystyle{4M_{N}\gamma \ p \over \pi \left( p^{2}+\gamma ^{2}\right) ^{2}}}%
\left( 1-\frac{{\bf q}^{2}\left( p^{2}+3\gamma ^{2}\right) }{6\left(
p^{2}+\gamma ^{2}\right) ^{2}}\right) \quad ,  \nonumber \\
F_{4} &=&%
{\displaystyle{1\  \over \pi }}%
\text{Im}\left[ B_{0}{}^{2}\left( A_{-1}^{(^{1}S_{0},nn)}(p)+\epsilon
A_{0}^{(^{1}S_{0},nn)}(p)+\epsilon ^{2}A_{1}^{(^{1}S_{0},nn)}(p)\right) %
\right] \ \quad .
\end{eqnarray}

\begin{eqnarray}
G_{2} &=&-\epsilon 
{\displaystyle{M_{N} \over 3\pi ^{2}}}%
\mathop{\rm Im}%
\left[ \sqrt{\frac{\gamma }{2\pi }}B_{0}C_{A}^{(1)}\widetilde{L}_{1,A}\left(
A_{-1}^{(^{1}S_{0},nn)}(p)+\epsilon A_{0}^{(^{1}S_{0},nn)}(p)\right) \right]
\nonumber \\
&&+\epsilon ^{2}%
{\displaystyle{M_{N}^{2}\gamma  \over 96\pi ^{4}}}%
\mathop{\rm Im}%
\left[ \widetilde{L}_{1,A}^{2}A_{-1}^{(^{1}S_{0},nn)}(p)\right] \quad , \\
G_{3} &=&\epsilon ^{2}%
{\displaystyle{M_{N} \over 3\pi ^{2}}}%
\mathop{\rm Im}%
\left[ \sqrt{\frac{\gamma }{2\pi }}B_{0}\left( C_{M}^{(1)}\widetilde{L}%
_{1,A}+C_{A}^{(1)}\widetilde{L}_{1,M}\ \right) A_{-1}^{(^{1}S_{0},nn)}(p)%
\right] \quad .  \nonumber
\end{eqnarray}
\begin{eqnarray}
\widetilde{L}_{1,A} &=&-\frac{4\pi \left( \mu -\gamma \right) }{%
M_{N}C_{0}^{(^{1}S_{0},nn)}}\left[ \sqrt{2}\left| V_{ud}\right| L_{1,A}-2\pi
C_{A}^{(1)}\left( \frac{M_{N}}{2\pi }C_{2}^{(^{1}S_{0},nn)}+\frac{\rho _{d}}{%
\left( \mu -\gamma \right) ^{2}}\right) \right] \quad ,  \nonumber \\
\widetilde{L}_{1,M} &=&-\frac{8\pi \left( \mu -\gamma \right) }{%
M_{N}C_{0}^{(^{1}S_{0},nn)}}\left[ \sqrt{2}\left| V_{ud}\right|
M_{N}L_{1}-\pi C_{M}^{(1)}\left( \frac{M_{N}}{2\pi }C_{2}^{(^{1}S_{0},nn)}+%
\frac{\rho _{d}}{\left( \mu -\gamma \right) ^{2}}\right) \right] \quad ,
\end{eqnarray}
evaluated at $\mu=m_\pi$, with $C_0^{(\,^1S_0,nn)}=-3.49$~fm$^2$ and 
$C_2^{(\,^1S_0,nn)}=6.46$~fm$^4$. 

\subsection{$\protect\nu_e +d\rightarrow e^-+p+p$}

\noindent
Differential cross section: 
\begin{eqnarray}
{\displaystyle{d^{2}\sigma  \over d\omega ^{^{\prime }}d\Omega }}%
&=&%
{\displaystyle{G_{F}^{2}\omega ^{\prime }\left| {\bf k}^{\prime }\right|  \over 2\pi ^{2}}}%
\frac{2\pi \eta _{e}}{1-e^{-2\pi \eta _{e}\eta _{e}}\ }\left[ 2W_{1}\sin ^{2}%
{\displaystyle{\theta  \over 2}}%
+W_{2}\cos ^{2}%
{\displaystyle{\theta  \over 2}}%
-2%
{\displaystyle{(\omega +\omega ^{\prime }) \over M_{d}}}%
W_{3}\sin ^{2}%
{\displaystyle{\theta  \over 2}}%
\right] \quad ,  \nonumber \\
\eta _{e} &=&\frac{2\alpha \omega ^{\prime }}{\left| {\bf k}^{\prime
}\right| }\quad .
\end{eqnarray}
Phase space:
\begin{gather}
\text{Max}\left[ -1,-{\displaystyle{\frac{4M_{N}(\nu -B+m_{n}-m_{p})-\omega
^{2}-\omega ^{\prime 2}+m_{e}^{2}}{2\omega \sqrt{\omega ^{\prime 2}-m_{e}^{2}%
}\ }}}\right] \leq \cos \theta \leq 1\quad ,  \nonumber \\
m_{e}\leq \omega ^{\prime }\leq \omega -2(M_{N}-\sqrt{%
M_{N}^{2}-M_{N}(B-m_{n}+m_{p})})
\end{gather}
Structure functions:
\begin{eqnarray}
W_{1} &=&\frac{1}{\left( 1-\epsilon \gamma \rho _{d}\right) }\left[
2C_{A}^{(1)^{2}}F_{1}-%
{\displaystyle{1 \over 3}}%
C_{A}^{(1)^{2}}F_{2}+\frac{4}{3}C_{A}^{(1)^{2}}F_{4}+G_{2}\right] \quad , 
\nonumber \\
W_{2} &=&W_{1}+\frac{1}{\left( 1-\epsilon \gamma \rho _{d}\right) }\left[
2C_{V}^{(1)^{2}}F_{1}-C_{V}^{(1)^{2}}F_{2}\right] \quad ,  \nonumber \\
W_{3} &=&\frac{-2\epsilon }{\left( 1-\epsilon \gamma \rho _{d}\right) }\left[
2C_{A}^{(1)}C_{M}^{(1)}F_{1}-%
{\displaystyle{1 \over 3}}%
C_{A}^{(1)}C_{M}^{(1)}F_{2}+\frac{4}{3}C_{A}^{(1)}C_{M}^{(1)}F_{4}+G_{3}%
\right] \quad .
\end{eqnarray}
\begin{eqnarray}
F_{1} &=&%
{\displaystyle{2M_{N}\gamma \ p \over \pi \left( p^{2}+\gamma ^{2}\right) ^{2}}}%
\left( 1+\frac{{\bf q}^{2}\left( p^{2}-\gamma ^{2}-2\eta p\gamma \right) }{%
2\left( p^{2}+\gamma ^{2}\right) ^{2}}\right) \left( 
{\displaystyle{2\pi \eta  \over \ e^{2\pi \eta }-1\ }}%
\right) e^{4\eta \tan ^{-1}\left( \frac{p}{\gamma }\right) }\quad , 
\nonumber \\
\quad F_{2} &=&%
{\displaystyle{4M_{N}\gamma \ p \over \pi \left( p^{2}+\gamma ^{2}\right) ^{2}}}%
\left( 1-\frac{{\bf q}^{2}\left( p^{2}+3\gamma ^{2}+6\eta p\gamma \right) }{%
6\left( p^{2}+\gamma ^{2}\right) ^{2}}\right) \left( 
{\displaystyle{2\pi \eta  \over \ e^{2\pi \eta }-1\ }}%
\right) e^{4\eta \tan ^{-1}\left( \frac{p}{\gamma }\right) }\quad \quad , 
\nonumber \\
F_{4} &=&%
{\displaystyle{1\  \over \pi }}%
\text{Im}\left[ B_{0}{}^{2}\left( A_{-1}^{(^{1}S_{0},pp)}(p)+\epsilon
A_{0}^{(^{1}S_{0},pp)}(p)+\epsilon ^{2}A_{1}^{(^{1}S_{0},pp)}(p)\right) %
\right] \quad ,
\end{eqnarray}
\begin{eqnarray}
B_{0} &=&M_{N}\int \frac{d^{3}k}{\left( 2\pi \right) ^{3}}\frac{\sqrt{8\pi
\gamma }}{k^{2}+\gamma ^{2}}%
{\displaystyle{2\pi \eta _{k} \over e^{2\pi \eta _{k}}-1\ }}%
\frac{e^{2\eta _{k}\tan ^{-1}\left( \frac{k}{\gamma }\right) }}{%
p^{2}-k^{2}+i\epsilon }  \nonumber \\
&&\qquad \qquad \qquad \qquad \left[ 1+\frac{{\bf q}^{2}\left( \left(
1-2\eta _{k}^{2}\right) k^{2}-6\eta _{k}k\gamma -3\gamma ^{2}\right) }{%
12\left( q^{2}+\gamma ^{2}\right) ^{2}}\right] \quad ,  \nonumber \\
\eta &=&\frac{\alpha M_{N}}{2p}\quad ,\quad \eta _{k}=\frac{\alpha M_{N}}{2k}%
\quad .
\end{eqnarray}
\begin{equation}
p=\sqrt{M_{N}\left( \nu -m_{n}+m_{p}\right) -\gamma ^{2}-\frac{{\bf q}^{2}}{4%
}+i\epsilon }\quad ,
\end{equation}
\[
C_{V}^{(1)}=%
{\displaystyle{\left| V_{ud}\right|  \over \sqrt{2}}}%
\quad ,\quad C_{A}^{(1)}=%
{\displaystyle{\left| V_{ud}\right|  \over \sqrt{2}}}%
g_{A}\quad ,\quad C_{M}^{(1)}=\sqrt{2}\left| V_{ud}\right| \kappa
^{(1)}\quad 
\]

\begin{eqnarray}
G_{2} &=&-\epsilon 
{\displaystyle{M_{N} \over 3\pi ^{2}}}%
\mathop{\rm Im}%
\left[ \sqrt{\frac{\gamma }{2\pi }}B_{0}C_{A}^{(1)}\widetilde{L}_{1,A}\left(
A_{-1}^{(^{1}S_{0},pp)}(p)+\epsilon A_{0}^{(^{1}S_{0},pp)}(p)\right) \right]
\nonumber \\
&&+\epsilon ^{2}%
{\displaystyle{M_{N}^{2}\gamma  \over 96\pi ^{4}}}%
\mathop{\rm Im}%
\left[ \widetilde{L}_{1,A}^{2}A_{-1}^{(^{1}S_{0},pp)}(p)\right] \quad , \\
G_{3} &=&\epsilon ^{2}%
{\displaystyle{M_{N} \over 3\pi ^{2}}}%
\mathop{\rm Im}%
\left[ \sqrt{\frac{\gamma }{2\pi }}B_{0}\left( C_{M}^{(1)}\widetilde{L}%
_{1,A}+C_{A}^{(1)}\widetilde{L}_{1,M}\ \right) A_{-1}^{(^{1}S_{0},pp)}(p)%
\right] \quad .  \nonumber
\end{eqnarray}
\begin{eqnarray}
\widetilde{L}_{1,A} &=&-\frac{4\pi \left( \mu -\gamma \right) }{%
M_{N}C_{0}^{(^{1}S_{0},pp)}}\left[ \sqrt{2}\left| V_{ud}\right| L_{1,A}-2\pi
C_{A}^{(1)}\left( \frac{M_{N}}{2\pi }C_{2}^{(^{1}S_{0},pp)}+\frac{\rho _{d}}{%
\left( \mu -\gamma \right) ^{2}}\right) \right] \quad ,  \nonumber \\
\widetilde{L}_{1,M} &=&-\frac{8\pi \left( \mu -\gamma \right) }{%
M_{N}C_{0}^{(^{1}S_{0},pp)}}\left[ \sqrt{2}\left| V_{ud}\right|
M_{N}L_{1}-\pi C_{M}^{(1)}\left( \frac{M_{N}}{2\pi }C_{2}^{(^{1}S_{0},pp)}+%
\frac{\rho _{d}}{\left( \mu -\gamma \right) ^{2}}\right) \right] \quad ,
\end{eqnarray}
evaluated at $\mu=m_\pi$, with $C_0^{(\,^1S_0,pp)}=-3.77$~fm$^2$ and 
$C_2^{(\,^1S_0,pp)}=7.50$~fm$^4$.

\end{document}